\documentstyle[11pt]{article}

\makeatletter
\@addtoreset{equation}{section}
\makeatother

\def\dalemb#1#2{{\vbox{\hrule height .#2pt
        \hbox{\vrule width.#2pt height#1pt \kern#1pt
                \vrule width.#2pt}
        \hrule height.#2pt}}}

\def\td{\tilde}

\let\a=\alpha \let\b=\beta   
    
 \let\m=\mu

\def\nn{\nonumber} \def\bd{\begin{document}} \def\ed{\end{document}}
\def\ds{\documentstyle} \let\fr=\frac \let\bl=\bigl \let\br=\bigr
\let\Br=\Bigr \let\Bl=\Bigl 
\let\bm=\bibitem
\let\na=\nabla
\let\pa=\partial \let\ov=\overline 
\newcommand{\be}{\begin{equation}} 
\newcommand{\ee}{\end{equation}} 
\def\ba{\begin{array}}
\def\ea{\end{array}}
\def\ft#1#2{{\textstyle{{\scriptstyle #1}\over {\scriptstyle #2}}}}
\def\fft#1#2{{#1 \over #2}}
\def\del{\partial}
\def\sst#1{{\scriptscriptstyle #1}}
\def\oneone{\rlap 1\mkern4mu{\rm l}}
\def\ie{{\it i.e.\ }}
\def\via{{\it via}}
\def\bog{{Bogomol'nyi\ }}
\def\vp{\varphi}
\def\im{{\rm i}}
\newcommand{\ho}[1]{$\, ^{#1}$}
\newcommand{\hoch}[1]{$\, ^{#1}$}
\newcommand{\bea}{\begin{eqnarray}} 
\newcommand{\eea}{\end{eqnarray}} 
\newcommand{\ra}{\rightarrow}
\newcommand{\lra}{\longrightarrow}
\newcommand{\Lra}{\Leftrightarrow}
\newcommand{\ap}{\alpha^\prime}
\newcommand{\bp}{\tilde \beta^\prime}
\newcommand{\tr}{{\rm tr} }
\newcommand{\Tr}{{\rm Tr} } 
\newcommand{\NP}{Nucl. Phys. }
\newcommand{\tamphys}{\it Center for Theoretical Physics,
Texas A\&M University, College Station, Texas 77843}
\newcommand{\ens}{\it Laboratoire de Physique Th\'eorique de l'\'Ecole
Normale Sup\'erieure\hoch{2}\\
24 Rue Lhomond - 75231 Paris CEDEX 05}

\newcommand{\auth}{H. L\"u\hoch{\dagger} and
C.N. Pope\hoch{\ddagger1}}

\begin{document}
\begin{flushright}
\hfill{CTP TAMU-11/97}\\
\hfill{LPTENS-97/06}\\
\hfill{hep-th/9702086}\\
\hfill{Feb. 1997}\\
\end{flushright}

\vspace{20pt}

\begin{center}
{ \large {\bf $p$-brane Taxonomy}}\hoch{*}

\vspace{30pt}
\auth

\vspace{15pt}

{\hoch{\dagger}\ens}

\vspace{10pt}
{\hoch{\ddagger}\tamphys}

\vspace{40pt}

\underline{ABSTRACT}
\end{center}

     We review an approach to the construction and classification of
$p$-brane solitons in arbitrary dimensions, with an emphasis on those that
arise in toroidally-compactified M-theory.  Procedures for constructing
the low-energy supergravity limits in arbitrary dimensions, and for
studying the supersymmetry properties of the solitons are presented.  Wide
classes of $p$-brane solutions are obtained, and their properties and
classification in terms of bound states and intersections of M-branes are
described.

{\vfill\leftline{}\vfill
\vskip	10pt
\footnoterule
{\footnotesize \hoch{*} Based on lectures presented at the Summer School in
High-Energy Physics and Cosmology, Trieste, \phantom{abcde}
Italy, 10 Jun - 26 Jul 1996. }\vskip 10pt
 {\footnotesize	\hoch{1} Research
supported in part by DOE  Grant DE-FG05-91-ER40633 \vskip -12pt} \vskip 10pt
{\footnotesize
        \hoch{2} Unit\'e Propre du Centre National de la Recherche
Scientifique, associ\'ee \`a l'\'Ecole Normale Sup\'erieure 
\phantom{abcde} et \`a l'Universit\'e de Paris-Sud}} 

\pagebreak
\setcounter{page}{1}

\section{Introduction}

     String theory and its goals have undergone a number of dramatic
re-appraisals since it was first introduced.  First seen as a
candidate for describing the strong interactions \cite{nam}, it fell into
disfavour on account of the fact that its spectrum included a massless
spin-2 excitation that was not seen in the hadronic arena.  A subsequent
revival of interest resulted from the realisation that this spin-2 state
should be interpreted not as a hadron but as the graviton \cite{yon,scsc}.  
With
this change of emphasis, string theory moved to the forefront of attempts
to find a framework for describing a quantum theory of gravity, and even
more ambitiously, a unified ``theory of everything."  Encouraged
by the discovery that considerations of anomaly-freedom greatly
restricted the possible gauge groups \cite{grsc}, and by the subsequent
discovery of the heterotic string, many attempts were to make contact
with the phenomenological world, of relatively low energy compared with
the Planck-scale unification regime.  It is probably fair to say
that early claims of the virtual uniqueness and predictive power of the
passage to the phenomenological arena have proved to be an exaggeration,
and at present the best that can be said is that at least there seem to be
ways of embedding the standard model into the theory.  After a fallow
period during which little further progress was achieved, the subject
was revolutionised again in 1995 with far-reaching discoveries about some
non-perturbative aspects of string theory.  Most notable of these was
the observation that by including non-perturbative states in the
spectrum of the type IIA string, whose presence is indicated by duality
symmetries, the degrees of freedom would be described in the
strong-coupling regime not by a ten-dimensional theory, but instead an
eleven-dimensional one \cite{ht,w1}.  In fact this discovery has led to a
revival of the fortunes of eleven-dimensional supergravity \cite{cjs},
which, although extensively investigated in the past as a possible
candidate for superunification (see, for example, \cite{dnp}), had long ago
been abandoned on account of its apparent inability to yield a realistic
four-dimensional low-energy description of the world.  It is now viewed as
the low-energy  limit of some yet to be discovered M-theory, which would
provide the more appropriate description of the degrees of freedom of the
type IIA string in all except the perturbative weak-coupling regime
\cite{w1}.  

      Evidence for the ability of $D=11$ supergravity to describe
elements of the spectrum of the type IIA string can be seen by
considering the BPS-saturated $p$-brane soliton solutions in the
low-energy effective theory from the type IIA string.  Included in
these are not only electrically-charged string solutions, which can be
directly identified with elementary string states in the perturbative
spectrum, but also other solutions that are more akin to solitons.  The
conjectured duality symmetries of the string, together with the fact
that these solutions are BPS saturated, and thus are expected to be
protected at the quantum level, lead to the expectation that they can be
identified with non-perturbative quantum states in the string spectrum. 
There are, for example, BPS-saturated black hole solutions in the type
IIA low-energy effective theory whose mass spectrum can be shown to
coincide precisely with the spectrum of massive particles coming from
the Kaluza-Klein dimensional reduction of $D=11$ supergravity on a
circle \cite{w1}.  

    The above, and other, considerations lead to a strong belief that
M-theory, and its low-energy $D=11$ supergravity limit, are relevant for
investigating the type IIA string.  A further surprise occurred when it
was argued that M-theory is also relevant for describing the $E_8\times
E_8$ heterotic string, by making the relatively
innocuous-sounding modification of compactifying it on $S^1/Z_2$ rather
than $S^1$ \cite{hw}.  

     With these, and other recent developments, the vital r\^ole of
M-theory and its low-energy $D=11$ supergravity limit have become
established.  In what follows, we shall present an overview of a
particular approach to the study of the solitonic $p$-brane spectrum of
the theory, and its toroidal compactifications to dimensions $D\le 11$.
Our emphasis will be on adopting a general and unified approach, in
which solutions in all dimensions can be studied and classified in a
systematic way.  We are aided in this aim by the fact that it is not
necessary, for these purposes, to have the complete
toroidally-compactified maximal supergravity theories at our disposal. 
The reason for this is that the $p$-brane solitons of interest here are
purely bosonic solutions, and the fermionic sectors of the theories need
be considered only insofar as they determine the fractions of
supersymmetry that are preserved by the solutions.  It turns out that
such questions can be answered very straightforwardly by reformulating
the problem as a bosonic one in $D=11$ itself, in a way that can then
easily be re-expressed in the lower-dimensional reduced theories without
the necessity of explicitly performing a dimensional reduction of the
fermionic sector of $D=11$ supergravity.

     We begin this review in section 2 by giving an explicit
construction of the bosonic sectors all the maximal supergravities in
$D\le 10$ that are obtained by toroidal compactification of $D=11$
supergravity.  In section 3, we discuss the basic structure of $p$-brane
soliton solutions, and we also introduce the \bog matrix in $D=11$,
and its dimensional reductions, which can be used to determine the
fractions of supersymmetry that are preserved by the various solitons.
In section 4, we discuss in detail various classes of $p$-brane solutions
in the maximal supergravities in arbitrary dimensions.  These include
extremal solutions, which saturate \bog bounds, and non-extremal solutions,
where the mass in general exceeds the \bog bound.  Our discussion includes
not only the standard kinds of $p$-brane solutions, but also a rather
general analysis of the equations of motion, yield additional solutions
that have received less attention in the literature.  Included in these are
different kinds of non-extremal $p$-brane solutions, and also non-extremal
$p$-branes that arise as solutions of certain systems of Toda equations.
Section 4 also includes a discssion of an interpretation for certain kinds
of $p$-branes as bound states of more fundamental ones.  Finally, in
section 5, we discuss the dimensional reduction and oxidation of $p$-brane
solutions.  Topics considered here include the two kinds of dimensional
reduction, corresponding to vertical and diagonal descent in a plot of
spacetime dimension versus $p$-brane dimension, and the oxidation of
lower-dimensional $p$-brane back to eleven dimensions, where some of them
acquire a new interpretation as intersections of the fundamental M-branes
of M-theory.

\section{Maximal supergravities in $D\le 11$}

     In this section, we discuss the toroidal dimensional reduction of
the bosonic sector of $D=11$ supergravity, whose Lagrangian takes the
form \cite{cjs}
\be
{\cal L} = \hat e \hat R -\ft1{48} \hat e\, \hat F_4^2 +\ft16 \hat F_4\wedge
\hat F_4\wedge \hat A_3
\ .\label{d11lag}
\ee
For brevity, we have written the final term as an 11-form; it is
understood that it should be dualised before integrating the Lagrangian
over the $D=11$ spacetime.  The subscripts on the potential $A_3$ and its
field strength $F_4=dA_3$ indicate the degrees of the differential forms.
We shall reduce the theory to $D$ dimensions in a succession of 1-step
compactifications on circles.  At each stage in the reduction, say from
$(D+1)$ to $D$ dimensions, the metric is reduced according to the standard
Kaluza-Klein prescription
\be
ds_{\sst D+1}^2 = e^{2\a\varphi} \, ds_{\sst D}^2 + e^{-2(D-2)\a\varphi}\, 
(dz+{\cal A}_1)^2\ ,\label{metred}
\ee
where the $D$ dimensional metric, the Kaluza-Klein vector potential ${\cal
A}_1={\cal A}_{\sst M}\, dx^{\sst M}$ and the dilatonic scalar $\varphi$
are taken to be independent of the additional coordinate $z$ on the
compactifying circle.  The constant $\a$ is given by
$\a^{-2} = 2(D-1)(D-2)$, and the parameterisation of the metric is such
that a pure Einstein action is reduced again to a pure Einstein action
together with canonically-normalised kinetic terms for ${\cal F}_2=d{\cal
A}_1$ and $\varphi$:
\be
e\, R \longrightarrow e\, R -\ft14 e\,
e^{-2(D-1)\a\varphi}\, {\cal F}_2^2 -\ft12 e\, (\del\varphi)^2\ .
\label{einstred}
\ee 

    Gauge potentials are reduced according to the prescription $A_n(x,z)=
A_n(x) + A_{n-1}(x)\wedge dz$, implying that a kinetic term for an $n$-form
field strength $F_n$ reduces according to the rule:
\be
-\fft1{2 \, n!}\, e \, F_n^2 \longrightarrow -\fft1{2 \, n!}\, e\,
e^{-2(n-1)\a\varphi}\,  F_n^2  -\fft1{2\, (n-1)!} \, e\,
e^{2(D-n)\a\varphi}\,  F_{n-1}^2\ .\label{fred}
\ee
There is a subtlety here in the expression for the dimensionally-reduced
field strength $F_n$, which is most easily seen by working in a vielbein
basis, since this facilitates the computation of the inner products in the
kinetic terms.  From the ansatz for the reduction of the gauge
potential we have
\be
F_n\longrightarrow dA_{n-1} + dA_{n-2}\wedge dz = dA_{n-1} -
dA_{n-2}\wedge {\cal A}_1 + dA_{n-2}\wedge (dz+{\cal A}_1)\ .\label{cs}
\ee
Thus while it is natural to define the dimensionally-reduced field
strength $F_{n-1}$ by $F_{n-1}=dA_{n-2}$, for $F_n$ we should define
$F_n=dA_{n-1} - dA_{n-2}\wedge {\cal A}_1$, and it is this
gauge-invariant field strength that appears on the right-hand side of
(\ref{fred}).  These so-called Chern-Simons modifications to the
lower-dimensional field strengths become progressively more complicated as
the descent through the dimensions continues. 

     It is not too difficult now to apply the above reduction procedures
iteratively, to construct the $D$-dimensional toroidally-compactified
theory from the $D=11$ starting point.  It is easy to see that the
original eleven-dimensional fields $g_{\sst{MN}}$ and $A_{\sst{MNP}}$ will
give rise to the following fields in $D$ dimensions,
\bea
g_{\sst{MN}} &\longrightarrow & g_{\sst{MN}}\ ,\qquad \vec\phi\ ,\qquad 
{\cal A}_1^{(i)}\ ,\qquad {\cal A}_0^{(ij)} \ ,\nn\\
A_3 &\longrightarrow & A_3\ ,\qquad A_2^{(i)}\ , \qquad A_1^{(ij)}\ ,
\qquad A_0^{(ijk)}\ ,\label{dfields}
\eea
where the indices $i, j, k$ run over the $11-D$
internal toroidally-compactified dimensions, starting from $i=1$ for the
step from $D=11$ to $D=10$.  The potentials $A_1^{(ij)}$ and 
$A_0^{(ijk)}$ are automatically antisymmetric in their internal indices,
whereas the 0-form potentials ${\cal A}_0^{(ij)}$ that come from the
subsequent dimensional reductions of the Kaluza-Klein vector potentials
${\cal A}_1^{(i)}$ are defined only for $j>i$.  The quantity $\vec\phi$
denotes the $(11-D)$-vector of dilatonic scalar fields coming from the
diagonal components of the internal metric.

     The Lagrangian for the bosonic $D$-dimensional toroidal
compactification of eleven-dimensional supergravity then takes the form
\cite{lpsol}
\bea
{\cal L} &=& eR -\ft12 e\, (\del\vec\phi)^2 -\ft1{48}e\, e^{\vec a\cdot 
\vec\phi}\, F_4^2 -\ft{1}{12} e\sum_i 
e^{\vec a_i\cdot \vec\phi}\, (F_3^{i})^2
-\ft14 e\, \sum_{i<j} e^{\vec a_{ij}\cdot \vec\phi}\, (F_2^{ij})^2
\nonumber\\
&& -\ft14e\, \sum_i e^{\vec b_i\cdot \vec\phi}\, ({\cal F}_2^i)^2
-\ft12 e\, \sum_{i<j<k} e^{\vec a_{ijk} \cdot\vec \phi}\,
(F_1^{ijk})^2 -\ft12e\, \sum_{i<j} e^{\vec b_{ij}\cdot \vec\phi}\,
({\cal F}_1^{ij})^2 + {\cal L}_{\sst{FFA}}\ ,\label{dgenlag}
\eea
where the ``dilaton vectors'' $\vec a$, $\vec a_i$, $\vec a_{ij}$, 
$\vec a_{ijk}$,
$\vec b_i$, $\vec b_{ij}$ are constants that characterise the couplings of
the dilatonic scalars $\vec \phi$ to the various gauge fields.  
They are given by \cite{lpsol}
\vfill\eject
 
\bea
&&F_{\sst{MNPQ}}\qquad\qquad\qquad\qquad\qquad\qquad\qquad\qquad
{\rm vielbein}\nonumber\\
{\rm 4-form:}&&\vec a = -\vec g\ ,\nonumber\\
{\rm 3-forms:}&&\vec a_i = \vec f_i -\vec g \ ,\nonumber\\
{\rm 2-forms:}&& \vec a_{ij} = \vec f_i + \vec f_j - \vec g\ ,
\qquad\qquad\qquad\qquad\qquad \,\,\, \,\vec b_i = -\vec f_i \ ,
\label{dilatonvec}\\
{\rm 1-forms:}&&\vec a_{ijk} = \vec f_i + \vec f_j + \vec f_k -\vec g
\ ,\qquad\qquad\qquad\qquad\vec b_{ij} = -\vec f_i + \vec f_j\ ,\nonumber \\
{\rm 0-forms:}&& \vec a_{ijk\ell} =\vec f_i +\vec f_j+\vec f_k +\vec f_\ell
-\vec g \ ,\qquad\qquad\quad\ \  \vec b_{ijk}=-\vec f_i +\vec f_j +\vec f_k\ ,
\nonumber
\eea
where the vectors $\vec g$ and $\vec f_i$ have $(11-D)$ components
in $D$ dimensions, and are given by
\bea
\vec g &=&3 (s_1, s_2, \ldots, s_{11-\sst D})\ ,\nonumber\\
\vec f_i &=& \Big(\underbrace{0,0,\ldots, 0}_{i-1}, (10-i) s_i, s_{i+1},
s_{i+2}, \ldots, s_{11-\sst D}\Big)\ ,\label{gfdef}
\eea
where $s_i = \sqrt{2/((10-i)(9-i))}$.  It is easy to see that they satisfy
\be
\vec g \cdot \vec g = \ft{2(11-D)}{D-2}, \qquad
\vec g \cdot \vec f_i = \ft{6}{D-2}\ ,\qquad
\vec f_i \cdot \vec f_j = 2\delta_{ij} + \ft2{D-2}\ .\label{gfdot}
\ee
We have also included the dilaton vectors $\vec a_{ijk\ell}$ and $\vec
b_{ijk}$ for ``0-form field strengths'' in (\ref{dilatonvec}), although they
do not appear in (\ref{dgenlag}), because they fit into the same general
pattern and they do arise if more general kinds of reduction procedure are
carried out \cite{ss1,bdgpt,clpst,lpdomain,llp}.
     
     The field strengths are associated with the gauge potentials in the
obvious way; for example $F_4$ is the field strength for $A_3$, $F_3^{(i)}$
is the field strength for $A_2^{(i)}$, {\it etc}.  In general, the field
strengths appearing in the kinetic terms are not simply the exterior
derivatives of their associated potentials, but have Chern-Simons
corrections as well, as discussed above.  On the other hand the terms
included in
${\cal L}_{\sst{FFA}}$, which denotes the dimensional reduction of the
$F_4\wedge F_4\wedge A_3$ term in $D=11$, are expressed purely in terms
of the potentials and their exterior derivatives.  The complete details of
all the field strengths, in the notation we are using here, were obtained in 
\cite{lpsol}.
The field strengths are given by
\bea
F_4 &=& \td F_4 - \gamma^{ij} \td F_3^i\wedge {\cal A}_1^j -\ft12
\gamma^{ik}\gamma^{j\ell} \td F_2^{ij} \wedge {\cal A}_1^k\wedge
{\cal A}_1^\ell + \ft16 \gamma^{i\ell}\gamma^{jm}\gamma^{kn}
\td F_1^{ijk}\wedge {\cal A}_1^\ell \wedge {\cal A}_1^m \wedge
{\cal A}_1^n\ ,\nonumber\\
F_3^i &=& \gamma^{ji}\td F_3^j - \gamma^{ji}\gamma^{k\ell} \td F_2^{jk}
\wedge {\cal A}_1^\ell - \ft12 \gamma^{ji}\gamma^{km}\gamma^{\ell n}
\td F_1^{jk\ell}\wedge {\cal A}_1^m \wedge {\cal A}_1^n\ ,\nonumber\\
F_2^{ij} &=& \gamma^{ki}\gamma^{\ell j} \td F_2^{k\ell} -
\gamma^{ki} \gamma^{\ell j} \gamma^{mn} \td F_1^{k\ell m}\wedge
{\cal A}_1^n\ ,\label{A.6}\\
F_1^{ijk} &=& \gamma^{\ell i} \gamma^{mj} \gamma^{nk} \td F_1^{\ell mn}
\ ,\nonumber\\
{\cal F}_2^i &=& \td {\cal F}_2^i - \gamma^{jk} \td {\cal F}_1^{ij} \wedge
{\cal A}_1^k\ ,\nonumber\\
{\cal F}_1^{ij} &=& \gamma^{kj} \td {\cal F}_1^{ik}\ ,\nn
\eea
where the tilded quantities represent the unmodified pure exterior
derivatives of the corresponding potentials, and $\gamma^{ij}$ is defined
by
\be
\gamma^{ij}=[(1+{\cal A}_0)^{-1}]^{ij}=\delta^{ij} -{\cal A}_0^{ij} +
{\cal A}_0^{ik}\, {\cal A}_0^{kj} +\cdots\ .\label{gam}
\ee
Recalling that ${\cal A}_0^{ij}$ is defined only for $j>i$ (and vanishes
if $j\le i$), we see that the series terminates after a finite number of
terms.

    The term ${\cal L}_{\sst{FFA}}$ in (\ref{dgenlag}) is the dimensional 
reduction of the $\td F_4\wedge\td F_4\wedge A_3$ term in $D=11$, and is
given in lower dimensions by \cite{lpsol}
\bea
D=10: &&\ft12 \td F_4\wedge \td F_4 \wedge A_2\ ,\nonumber\\
D=9: &&\Big(-\ft14 \td F_4 \wedge \td F_4 \wedge A_1^{ij}-\ft12 \td F_3^i
\wedge \td F_3^j \wedge A_3\Big)\epsilon_{ij}\ ,\nonumber\\
D=8: && \Big(-\ft1{12} \td F_4\wedge \td F_4 A_0^{ijk} -\ft16 \td F_3^i\wedge
\td F_3^j \wedge A_2^k +\ft12 \td F_3^i \wedge \td F_2^{jk} \wedge
A_3\Big) \epsilon_{ijk}\ ,\nonumber\\
D=7: && \Big(-\ft16 \td F_4\wedge \td F_3^i A_0^{jkl} +\ft16 \td F_3^{i}\wedge
\td F_3^{j} \wedge A_1^{kl} +\ft18 \td F_2^{ij}\wedge \td F_2^{kl}
\wedge A_3\Big)\epsilon_{ijkl}\ ,\label{ffaterms}\\
D=6: && \Big(\ft1{12} \td F_4\wedge \td F_2^{ij} A_0^{klm} +\ft1{12}
\td F_3^i\wedge \td F_3^j A_0^{klm} +\ft18 \td F_2^{ij}\wedge
\td F_2^{kl} \wedge A_2^m\Big) \epsilon_{ijklm}\ ,\nonumber\\
D=5: && \Big(\ft1{12} \td F_3^i\wedge \td F_2^{jk}  A_0^{lmn} -\ft1{48}
\td F_2^{ij}  \wedge \td F_2^{kl}\wedge A_1^{mn} -\ft1{72}
\td F_1^{ijk}\wedge \td F_1^{lmn} \wedge A_3\Big)
\epsilon_{ijklmn}\ ,\nonumber\\
D=4: && \Big(-\ft1{48} \td F_2^{ij}\wedge \td F_2^{kl} A_0^{mnp} -\ft1{72}
\td F_1^{ijk}\wedge \td F_1^{lmn} \wedge A_2^p\Big)
\epsilon_{ijklmnp}\ ,\nonumber\\
D=3:&& \ft1{144}\, \td F_1^{ijk}\wedge \td F_1^{lmn}\wedge A_1^{pq}
\epsilon_{ijklmnpq}\ ,\nonumber\\
D=2: && \ft1{1296}\, \td F_1^{ijk}\wedge \td F_1^{lmn} A_0^{pqr}
\epsilon_{ijklmnpqr}\ .\nonumber
\eea

     In the subsequent sections, we shall be making extensive use of the
results presented here, in order to discuss various aspects of $p$-brane
solitons in toroidally-compactified type II strings.

\section{$p$-branes and supersymmetry}

    Extremal $p$-branes in various supergravities in different dimensions
were constructed in the past [16-25].  Our principle aim in this
section will be to explain a procedure for determining the fractions
of supersymmetry that are preserved by the various $p$-brane solitons
that we shall be discussing later.  In order to set the stage for
this, it is necessary first for us to describe the basic structure of
the $p$-brane solitons.  They arise as solutions to the supergravity
theories described by (\ref{dgenlag}), where in any given solution
only a subset of the bosonic fields will be involved.  More
specifically, in a $p$-brane soliton solution the metric tensor, one
or more of the dilatonic scalars, and one or more field strengths are
active, where the degrees of the field strengths are either $(p+2)$ or
$(D-p-2)$.  In the former case, the field strengths carry
electric-type charges, whilst in the latter, they carry magnetic-type
charges.  The form of the metric is
\be
ds^2=e^{2A}\, dx^\mu\, dx^\nu\, \eta_{\mu\nu} + e^{2B}\, dy^m\, dy^m\ ,
\label{metricans1}
\ee
where $x^\mu$ are the coordinates on the world-volume for the $p$-brane,
of dimension $d=p+1$, and $y^m$ are the remaining $(D-d)$ coordinates of
the $D$-dimensional spacetime, which are transverse to the $p$-brane
worldsheet.  It will be convenient for future reference to define the
quantity $\td d=D-d-2$.    The functions $A$ and $B$ are independent of
the world-volume coordinates $x^\mu$.  In the simplest situation, where
one considers a single-centre $p$-brane solution which can be located at
the origin $y^m=0$ without loss of generality, $A$ and $B$ will depend
only on $r=\sqrt{y^m\, y^m}$.  These solutions will be sufficient for our
present discussion.  In this case, we may rewrite the ansatz
(\ref{metricans1}) using hyperspherical polar coordinates in the transverse
space thus \cite{dghr}:
\be
ds^2=e^{2A}\, dx^\mu\, dx^\nu\, \eta_{\mu\nu} + e^{2B}\, (dr^2+ r^2\,
d\Omega^2)\ ,
\label{metricans2}
\ee
where $d\Omega^2$ is the metric on the unit $(\td d+1)$-sphere.  In all
cases, the $p$-brane solutions are such that at large distance the metric
approaches the flat metric, the scalars become constant, and the field
strengths go to zero.

     The charges carried by the field strengths are measured by performing
appropriate surface integrals over the $(\td d+1)$-sphere at infinity.  If
a field strength $F$ carries electric charge $u$, or magnetic charge $v$,
then these are given by \cite{page}
\be
u=\fft{1}{4\omega_{\td d+1}}\int_{S^{\td d+1}}\, *F\ ,\qquad 
{\rm or}\qquad v=\fft{1}{4\omega_{\td d+1}}\int_{S^{\td d+1}}\, F\ ,
\label{emcharge}
\ee
respectively, where $\omega_{\td d+1}$ is the volume of the unit $(\td
d+1)$-sphere.  (We are assuming here for simplicity that the dilatonic
scalars, which are asymptotically constant at infinity, are chosen to vanish
there.)

     We are now ready to present a framework for discussing the
supersymmetry of the solutions.  We shall do this by first describing the
situation in $D=11$, and then performing a dimensional reduction to $D$
dimensions.  If an asymptotically-flat solution preserves some fraction of
the supersymmetry, there will exist Killing spinors $\epsilon$ that become
asymptotically constant at infinity.  From these, global supercharges can
be defined.  In $D=11$, this supercharge will be given by
\be
Q_\epsilon = \int_{\del\Sigma_{\td d+1}}\, \bar\epsilon\,
\Gamma^{\sst{MNP}}\,\psi_{\sst P}\, 
d\Sigma_{\sst{MN}}\ ,\label{supercharge}
\ee
where $\del\Sigma_{\td d+1}$ is the $(\td d+1)$-sphere of radius $r$ in
the transverse space.  The anti-commutator of the resulting supercharges is
given by
\be
\{ Q_{\epsilon_1}, \, Q_{\epsilon_2} \} = \delta_{\epsilon_1}\,
Q_{\epsilon_2} =\int_{\del \Sigma} N^{\sst{AB}}\, d\Sigma_{\sst{AB}}\ ,
\label{commutator}
\ee
where $N^{\sst{AB}} = \bar \epsilon_1\Gamma^{\sst{ABC}}\delta_{\epsilon_2}
\psi_{\sst C}$.   From the transformation rule for the gravitino in $D=11$
supergravity, we therefore obtain the Nester form
\be
N^{\sst{AB}} = \bar \epsilon_1 \Gamma^{\sst{ABC}}\, D_{\sst C} \epsilon_2 +
\ft18 \bar \epsilon_1 \Gamma^{\sst{C_1C_2}}\epsilon_2\,\,
F^{\sst{AB}}{}_{\sst{C_1C_2}} + \ft1{96} \bar \epsilon_1
\Gamma^{\sst{ABC_1\ldots C_4}}\epsilon_2\,\, F_{\sst{C_1\ldots C_4}}\ .
\label{nestor}
\ee
Since only the $d\Sigma_{0r}$ component of the $p$-brane spatial volume
element contributes in (\ref{commutator}), we may read off the Bogomol'nyi
matrix ${\cal M}$ from the integral
\be
 {1\over \omega_{\tilde d+1}}   \int_{\del\Sigma\, {\rm at}\,
r\rightarrow \infty} N^{0r} r^{\tilde d +1} d\Omega_{(\tilde d+1)}
 = \epsilon_1^\dagger {\cal M} \epsilon_2\ ,
\label{matdef}
\ee
where $\omega_{\tilde d + 1}$ is the volume of the unit $(\tilde d +1)$-sphere.
If there is an unbroken supersymmetry, then there exists a Killing spinor
such that eqn.\ (\ref{commutator}) vanishes.  In other words, the
Bogomol'nyi matrix (\ref{matdef}) has a zero eigenvalue for each component
of the unbroken supersymmetry.

    We can now use the Bogomol'nyi matrix to study the supersymmetry of
the $p$-brane solutions in $D=11$ dimensions.  There is only one field
strength in $D=11$ supergravity, namely the 4-form, which gives rise to an
electrically-charged membrane or a magnetically-charged 5-brane.  Note that
in the electric case the
last term in (\ref{nestor}) vanishes, whilst the second term vanishes in
the magnetic case.  Substituting (\ref{matdef}), we obtain \cite{dlps}
\bea
{\rm electric:} && {\cal M} = m\oneone + u \Gamma_{012}\ ,\nonumber\\
{\rm magnetic:} && {\cal M} = m\oneone + v \Gamma_{{\hat 1}{\hat 2}
{\hat 3}{\hat 4}{\hat 5}}\ ,\label{bog11}
\eea
where the hats indicate index values in the transverse space, while indices
without hats live in the world-brane volume, and $u$ and $v$ denote 
the electric and magnetic charges defined in (\ref{emcharge}).  The 
parameter $m$ denotes the mass per unit volume of the $p$-brane, which is
calculated using the ADM mass formula.  It is a measure of the rate at
which the metric approaches flatness at infinity, and arises from the
spin connection in the first term in (\ref{nestor}).  We shall postpone a
detailed discussion of the supersymmetry of specific solutions until
subsequent sections, but we just remark for now that one can easily
determine from (\ref{bog11}) that the eigenvalues of the \bog matrix
${\cal M}$ are given by $m\pm u$ or $m\pm v$, with
sixteen eigenvalues for each sign choice, and thus half the
supersymmetries are preserved if $u=m$ or $v=m$.  These correspond to the
BPS-saturated membrane or 5-brane in $D=11$.

     The above analysis of supersymmetry can easily be generalised to lower
dimensions.  In fact the Nester form for maximal supergravity in any
dimension is just the Kaluza-Klein dimensional reduction of the
11-dimensional expression (\ref{nestor}).  For example, the Nester form
upon reduction to type IIA supergravity in $D=10$ is given by
\cite{dghr,lpsol}
\bea
N^{\sst{AB}} &=& \bar \epsilon_1\Gamma^{\sst{ABC}} D_{\sst{C}} \epsilon_2 +
e^{-\ft34\phi}\, \bar \epsilon_1\Gamma_{10} \Big (\ft14 {\cal F}^{\sst{AB}} +
\ft18 \Gamma^{\sst{ABCD}} {\cal F}_{\sst{CD}} \Big)\epsilon_2\nonumber\\
&& - e^{\ft12\phi}\,\bar \epsilon_1\Gamma_{10}\Big( \ft1{4}
\Gamma^{\sst{C}} F^{\sst{AB}}
{}_{\sst C} +\ft1{24} \Gamma^{\sst{ABCDE}}
F_{\sst{CDE}}\Big)\epsilon_2\label{bog10all}\\
&& + e^{-\ft14 \phi}\, \bar \epsilon_1
\Big( \ft18 \Gamma^{\sst{CD}} F^{\sst{AB}}{}_{
\sst{CD}} + \ft1{96} \Gamma^{\sst{ABCDEF}} F_{\sst{CDEF}}\Big)
\epsilon_2\ .\nonumber
\eea
The Nester forms become increasingly complicated as we descend through the
dimensions, since more and more antisymmetric tensors are generated.  However,
for the purpose of studying the supersymmetries of $p$-brane solutions, some
simplifications can be made.   First, note that the dilaton factor for each
field strength is precisely the square root of the dilaton factor for the
kinetic term of the same field strength that appears in the Lagrangian.  In
fact all these dilaton factors can be
set to unity since the Bogomol'nyi matrix we are considering is defined at
$r=\infty$, and we are taking the dilatons to vanish there.  As we showed
above, in order to obtain the eigenvalues of a Bogomol'nyi matrix, we do not
need to decompose the
$\Gamma$ matrices into world-volume and transverse space factors. 
Furthermore, we do not need to decompose the 11-dimensional $\Gamma$ matrices
into the product of
$D$-dimensional spacetime and compactified $(11-D)$-dimensional factors.
This greatly simplifies the discussion for lower dimensions.

     In order to present the general Bogomol'nyi matrix for arbitrary forms
and arbitrary dimensions, we first establish a notation for the charges
carried by the various field strengths:
\bea
{}&&F_4\qquad\quad F_3^i\quad\qquad F_2^{ij}\qquad\quad F^{ijk}_1
\quad\qquad {\cal F}_2^i\qquad\quad {\cal F}_1^{ij}\nonumber\\
{\rm electric:}&& u\quad\qquad\,\,\, u_i\quad\qquad\,\, u_{ij}\quad\qquad
u_{ijk}\quad\qquad\,\, p_i\quad\qquad
p_{ij}\label{pagecharge}\\
{\rm magnetic:}&& v\qquad\quad \,\,\, v_i\quad\qquad \,\, v_{ij}\quad\qquad
\,\, v_{ijk}\quad\qquad \, q_i\quad\qquad\,\,\,\, q_{ij}\ ,\nonumber
\eea
where the electric $u$-type or $p$-type charges, and the magnetic $v$-type
or $q$-type charges, are given by (\ref{emcharge}).  We then find that the
general Bogomol'nyi matrix in $D$ dimensions is given by \cite{lpsol}
\bea
{\cal M} &=& m\oneone + u\, \Gamma_{012} + u_i\, \Gamma_{01i} +
\ft12 u_{ij}\, \Gamma_{0ij}+ \ft16 u_{ijk}\, \Gamma_{ijk} +p_i
\Gamma_{0i}+ \ft12 p_{ij}\, \Gamma_{ij} \label{genbog}\\ 
&&+v\,\Gamma_{\hat1\hat2\hat3\hat4\hat5} + v_i\,
\Gamma_{\hat1\hat2\hat3\hat4i}+\ft12 v_{ij}\, \Gamma_{\hat1\hat2\hat3ij}
+ \ft16 v_{ijk}\, \Gamma_{\hat1\hat2ijk} + q_i\, \Gamma_{\hat1\hat2\hat3 i}
+\ft12 q_{ij}\, \Gamma_{\hat1\hat2ij}\ ,\nn
\eea
where the first line contains the contributions for electrically-charged
solutions, and the second line contains the contributions for
magnetically-charged solutions.  For a given degree $n$ of antisymmetric
tensor field strength, only the terms with the corresponding charges, as
indicated in (\ref{pagecharge}), will occur.  As always, the indices $0,
1,\ldots$ run over the dimension of the $p$-brane worldvolume,
$\hat1,\hat2,\ldots$ run over the transverse space of the $y^m$ coordinates,
and $i,j,\ldots$ run over the dimensions that were compactified in the
Kaluza-Klein reduction from 11 to $D$ dimensions.  The mass per unit
$p$-volume $m$ in (\ref{genbog}) arises from the connection term in the
covariant derivative in the Nester form, and it is given by $m =\ft12 {\rm
lim}_{r \rightarrow \infty}\, (B'-A')e^{-B} r^{\tilde d + 1}$.

     In the subsequent sections, we 
shall make use of the \bog matrix constructed above in order to determine 
the fractions of supersymmetry that are preserved by the various $p$-brane
solutions.  

\section{$p$-brane solitons in maximal supergravities}

     When solving the equations of motion (\ref{dgenlag}) for $p$-brane
solutions with a given $p$, only the subset of field strengths whose
degrees are either $(p+2)$ (in the case of electric charges) or $(D-p-2)$
(in the case of magnetic charges) are involved.   Thus the relevant part of
the supergravity Lagrangian that describes the $p$-brane solutions will be
of the form
\be
{\cal L} = eR -\ft12 e\, (\del \vec\phi)^2 -\fft1{2n!}
\sum_{\alpha =1}^{N} e^{\vec c_{\alpha}\cdot \vec\phi}\, F_\alpha^2
\ ,\label{nlag}
\ee
where we suppose that $N$ field strengths $F_\a$ of
degree $(p+2)$ or $(D-p-2)$, labelled by $\a$, are active.  These field
strengths, and their associated dilaton vectors $\vec c_\a$, are therefore a
subset of the ones appearing in (\ref{dgenlag}).

\subsection{Multi-charge extremal solutions}

    We begin our discussion of $p$-brane solitons by considering the case
of extremal solutions.  We shall make the spherically-symmetric ansatz
(\ref{metricans2}) for the metric, while each field strength, carrying an
electric or a magnetic charge, will take the form
\be
F^\a_{m\mu_1\cdots\mu_{n-1}} = \epsilon_{\mu_1\cdots\mu_{n-1}} (e^{C_\a})'\,
\fft{y^m}{r}\qquad {\rm or}\qquad
F^\a_{m_1\cdots m_n }= \lambda_\a \,\epsilon_{m_1\cdots m_n p}\, 
\fft{y^p}{r^{n+1}}\ ,\label{fansatz}
\ee
where a prime denotes a derivative with respect to $r$.  These two
ans\"atze both preserve the same $SO(1,d-1)\times SO(D-d)$ subgroup of the 
original $SO(1,D-1)$ Lorentz group as does the metric (\ref{metricans2}).
Substituting the ans\"atze (\ref{fansatz}) and (\ref{metricans2}) directly
into the equations of motion that follow from the Lagrangian (\ref{nlag}),
we find that $\vec \phi$, $A$ and $B$ satisfy 
\vfill\eject

\bea
&&\vec \phi'' +\fft{\td d +1}{r}\, \vec \phi' +(dA' + \td d B')\vec \phi'
= -\ft12\epsilon \sum_\a \vec c_\a\, S^2_\a\ ,\label{eom11}\nn\\
&&A'' + \fft{\td d +1}{r}\, A' + (dA' + \td d B') A' =
\fft{\td d}{2(D-2)} \sum_\a S_\a^2\ ,\label{eom12}\nn\\
&&B'' +\fft{\td d +1}{r}\, B' + (dA' + \td d B') (B' + \fft1{r})
=-\fft{d}{2(D-2)} \sum_\a S_\a^2\ ,\label{eom13}\\
&&d(D-2) A'^2 + \td d (d A'' + \td d B'') - (d A' + \td d B')^2
-\fft{\td d}{r} (d A' + \td d B') + \ft12 \td d \, \vec\phi'^2 
= \ft12 \td d \sum_\a S_\a^2\ ,\nn
\eea
where $\epsilon =1$ or $-1$ for the electric or magnetic ansatz 
respectively, and the functions $S_\a$ are given by
\be
S_\a = \lambda_\a\, e^{-\ft12\epsilon \vec c_\a\cdot \vec \phi 
-\td d B}\, r^{-\td d -1}\ .\label{salpha}
\ee
In the electric case, $\lambda_\a$ arises as the integration constant for 
the function $C_\a$, given by
\be
(e^{C_\a})' = \lambda_\a \, e^{\vec c_\a \cdot \vec \phi + d A - \td B}\,
r^{-\td d -1}\ . 
\ee

      From (\ref{eom12}) and (\ref{eom13}), we see that a natural solution 
for $B$ is to take 
\be
d A + \td d B=0\ .\label{dab}
\ee 
(We shall return later to the discussion of more general solutions in
which this relation is not imposed.)  We may also consistently set to zero
the $(11-D-N)$ components of $\vec \phi$ that are 
orthogonal to the space spanned by the $N$ dilaton vectors $\vec c_\a$. 
The  remaining equations become
\bea
&&\varphi_\a'' + \fft{\td d +1}{r}\, \varphi_\a' =
-\ft12 \epsilon \sum_\beta M_{\a\beta}\, S^2_\beta\ ,\label{eom21}\\
&&A'' + \fft{\td d +1}{r} \, A' = \fft{\td d}{2(D-2)} \sum_\a S_\a^2
\ ,\label{eom22}\\
&&d(D-2) A'^2 +\ft12 \td d \sum_{\a,\beta} (M^{-1})_{\a\beta}\, \varphi_\a'
\varphi_\beta' =\ft12 \td d \sum_\a S_\a^2\ ,\label{eom23}
\eea
where we have defined $\varphi_\a = \vec c_\a \cdot \vec \phi$, and
$M_{\a\b}$ is the matrix of dot products of the dilaton vectors
\be
M_{\a\b}= \vec c_\a\cdot \vec c_\b\ .
\ee
(Here we are  assuming that
$M_{\a\beta}$ is non-singular, and we shall comment on the  case when it is
singular later.) Note that the number of non-vanishing scalar fields
$\varphi_\a$ is precisely the same as the number $N$ of participating field
strengths. By acting on (\ref{eom21}) with
$(M^{-1})_{\a\beta}$, and comparing with (\ref{eom22}), we see that it is
natural to solve for $A$ by taking 
\be
A =- \fft{\epsilon \td d}{D-2} 
\sum_{\a,\beta} (M^{-1})_{\a\beta}\, \varphi_\a\ .
\ee
The equations of motion now reduce to 
\bea
\sum_{\beta} (M^{-1})_{\a\beta}\Big(\varphi_\beta'' + 
\fft{\td d +1}{r}\, \varphi_\beta'\Big) 
&=& -\ft12 \epsilon \lambda^2_\a\, 
e^{-\epsilon\varphi_\a + 2d A}\, r^{-2(\td d +1)}\ ,
\label{eom31}\\
d(D-2) A'^2 + \ft12 \td d \sum_{\a,\beta} (M^{-1})_{\a\beta} \,
\varphi_\a'\varphi_\beta' &=&\ft12 \td d \sum_\a \lambda_\a^2 \,e^{-\epsilon
\varphi_\a + 2d A}\, r^{-2(\td d+1)}\ .\label{eom32}
\eea
The solutions are determined completely by the structure of the
dot products $M_{\a\beta}$ of dilaton vectors $\vec c_\a$ of the
corresponding field strengths $F^\a$.  Solutions exist only for $N\le
(11-D)$.  In general, the solutions of (\ref{eom31}) and (\ref{eom32}) are 
still very complicated.   However, we can find simple solutions if we make
the ansatz that the quantity $(-\epsilon \varphi_\a + 2d A)$ appearing in
the exponential in $S_\a^2$ is proportional to the quantity $\sum_{\beta}
(M^{-1})_{\a\beta}\, \varphi_\beta$ appearing on the left-hand side of
(\ref{eom31}).  For this to be true, it implies that $M_{\a\beta}$ must take
the form 
\be
M_{\a\beta} = 4 \delta_{\a\beta} - \fft{2d\td d}{D-2}\ .\label{mmatrix}
\ee
Note that the coefficient of $\delta_{\a\beta}$ can {\it a priori} be any
constant, but it is fixed to be 4 in maximal supergravity theories, as
can be verified by computing the magnitudes of all the dilaton vectors,
defined by (\ref{dilatonvec}).  We can now solve  (\ref{eom31}) and
(\ref{eom32}) completely by making the further ansatz that
$S_\a \propto (-\epsilon \varphi_\a' + 2d A')$. Thus the solutions for the
dilaton and $p$-brane metric are \cite{lpmulti}
\bea
&&e^{\ft12 \epsilon \varphi_\a - d A} = 1 + \fft{\lambda_\a}{\td d} r^{-\td 
d}\ ,\label{gensol1}\\
&&ds^2=\prod_{\a=1}^{N} \Big(1+ \fft{\lambda_\a}{\td d} 
r^{-\td d}\Big)^{-\ft{\tilde d}{(D-2)}}\, dx^\mu dx^\nu \eta_{\mu\nu} +
\prod_{\a=1}^N \Big(1+ \fft{\lambda_\a}{\td d} 
r^{-\td d}\Big)^{\ft{d}{(D-2)}}\, (dr^2 +r^2 d\Omega^2)
\ .\nn
\eea
Note that the functions $H_\a\equiv(1+ \ft{\lambda_\a}{\td d} r^{-\td d})$
are harmonic on the internal space, and thus we may express the solution
more succinctly in terms of these harmonic functions \cite{kklp,drbound}, 
\bea
 &&e^{\ft12 \epsilon \varphi_\a - d A} = H_\a\ ,\nonumber\\
&&ds^2=\prod_{\a=1}^{N} H_\a^{-\ft{\tilde d}{(D-2)}}\, dx^\mu dx^\nu
\eta_{\mu\nu} +
\prod_{\a=1}^N H_\a^{\ft{d}{(D-2)}}\, dy^m \, dy^m
\ .\label{gensol2}
\eea     
It is straightforward to see from the ans\"atze (\ref{fansatz}) that the
field strengths are given in the electric or magnetic cases by
\be
F_\a = dH^{-1}_\a\wedge d^dx\ ,\qquad {\rm or} \qquad 
F_\a =*(dH^{-1}_\a\wedge d^dx)\ ,\label{multfsol}
\ee
respectively.  The extremal metrics given in (\ref{gensol1}) have an
horizon at $r=0$.  In general, this coincides with a singularity of the
curvature tensor, and also the dilatonic scalars diverge there.  In the
special cases where the dilatons are finite on the horizon, the curvature
is finite there too.  (An example of this arises in the four-charge
solution in $D=4$, for which it can easily be verified that the dilatonic
scalars are finite on the horizon at $r=0$.  If the charges are set equal,
then the dilatonic scalars become constants everywhere, and in fact the
solution reduces to the extremal Reissner-Nordstr{\o}m solution.)  It should
be remarked also that if the matrix $M_{\a\b}$ defined by (\ref{mmatrix})
happens to be degenerate the solutions are still given by (\ref{gensol2}),
but now it turns out that some linear combination of the dilatonic scalars
$\varphi_\a$ vanishes, and so there is one fewer scalar degrees of freedom
in such cases.

    Solutions of the above kind can be found in the maximal supergravity
theories in each dimension $D\le 11$.  The values of $p$ for which
solutions exist are determined by the degrees of the fields strengths that
exist in the particular dimension $D$ in question.  The solutions in
general carry $N$ independent electric or magnetic charges, which, from 
(\ref{multfsol}) and (\ref{emcharge}), are easily found to be
given by $Q_\a= \ft14 \lambda_\a$. The mass per unit $p$-volume can also
easily be calculated, and turns out to be given by
\be
m=\sum_{\a=1}^N Q_\a\ .\label{massform}
\ee
The number $N$ of independent charges that can arise for a given $p$ in a
given dimension $D$ depends on two factors.  First of all, $N$ is
certainly bounded by the number of dilaton vectors $\vec c_\a$ in the
toroidally-compactified theory whose dot products satisfy the necessary
relation (\ref{mmatrix}).  If the field strengths appearing in
(\ref{dgenlag}) and (\ref{nlag}) were all simply the exterior derivatives
of their associated potentials, then in fact this  would be the only
criterion determining the numbers of field strengths that could be used in
constructing multi-charge solutions.  However, as we saw in the previous
section, there are Chern-Simons corrections in the expressions for the
field strengths, and these imply that the complete system of field
equations for the fields in the supergravity theories are much more
complicated than at first sight might appear.  In particular, Chern-Simons
corrections involving a field that is being set to zero in a particular
solution can nevertheless impose constraints on the fields that are
retained, since one must vary the Lagrangian with respect to all the
fields before setting any of them to zero.  The complete analysis of all
possible $p$-brane solutions is therefore extremely complicated.  In
practice, a useful strategy for approaching the problem is to proceed
first with finding configurations that would be solutions in the absence
of the Chern-Simons complications, and then check which of them survives
after taking account of the constraints implied by the setting to zero of
the non-participating fields.  It is not certain that one will find all
solutions by this means, but at least one will find some of them.  Indeed,
it is not clear to us that any completely exhaustive discussion of the
solution set has been given in the literature.

     The known multi-charge extremal solutions can be summarised as
follows. Using the 4-form field strength, we can clearly only construct
single-charge solutions, since there is only one such field strength.  If
it carries an electric charge, we obtain an extremal membrane solution
for each $D$, while if it carries a magnetic charge, we get an extremal
$(D-6)$-brane.  For 3-form field strengths, it turns out that although
there will more than one of them in each dimension $D\le 9$, their dilaton
vectors never satisfy the necessary relation (\ref{mmatrix}), and
consequently one can only obtain single-charge solutions.  Thus we have
single-charge extremal solutions in $D\le 10$, which are strings if the
charge is electric, and $(D-5)$-branes if the charge is magnetic.  For the
case of solutions using 2-form or 1-form field strengths, it turns out
that multi-charge solutions can arise.  The possibilities are summarised
in the following table \cite{lpsol}:

\bigskip

\centerline{
\begin{tabular}{|c||c|l||c|c|}\hline
{\phantom{D}Dim.\phantom{D}}&\multicolumn{2}{c||} 
{\phantom{DDDDD} 2-Forms \phantom{DDDDD}} & 
\multicolumn{2}{c|}{\phantom{DDDDD}1-Forms\phantom{DDDDD}} \\ \hline\hline
$D=10$& $\phantom{D}N=1\phantom{D}$ &\phantom{DD}$p=0,6$ & &  \\ \hline
$D=9$ & $N=2$ &\phantom{DD}$p=0,5$ & $N=1$ & $p=6$ \\ \hline
$D=8$ &       &\phantom{DD}$p=0,4$ &$N=2$&$p=5$ \\ \hline
$D=7$ &       &\phantom{DD}$p=0,3$ &     &$p=4$ \\ \hline
$D=6$ &       &\phantom{DD}$p=0,2$ &$N=3,4'$&$p=3$ \\ \hline
$D=5$ & $N=3$ &\phantom{DD}$p=0,1$ &        &$p=2$ \\ \hline
$D=4$ & $N=4$ &\phantom{DD}$p=0$   &$N=4,5,6,7$&$p=1$ \\ \hline
$D=3$ &       &                    &$N=8$      &$p=0$ \\ \hline
\end{tabular}}
\bigskip

\centerline{Table 1: Numbers of charges in multi-scalar $p$-brane solutions}
\bigskip

\noindent  Here we list the highest dimensions where $p$-brane solutions
with the indicated numbers $N$ of field strengths first occur.  They then
occur also at all lower dimensions.

     Special cases of the multi-charge solutions arise if all $N$ charges
are set equal, in which case the harmonic functions $H_\a$ in
(\ref{gensol2}) become equal.  Under these circumstances, it is easy to
see from (\ref{gensol2}) that all except one combination of the dilatonic
scalar will become zero, and at the same time all the participating field
strengths will become equal.  The resulting single-scalar configuration is 
a solution of the truncated Lagrangian
\be
{\cal L} = e
R -\ft12 e\, (\del\phi)^2 -\fft1{2n!} e\, e^{a\phi} F^2 \ ,\label{slag}
\ee
and is given by \cite{lpss1}
\bea
e^{\ft{\epsilon\Delta}{2a}\phi}&=& H\ ,\nn\\
ds^2 &=& H^{-\ft{4\td d}{\Delta(D-2)}}\, dx^\m\, dx^\nu\, \eta_{\mu\nu} +
H^{\ft{4 d}{\Delta(D-2)}}\,(dr^2 + r^2 \, d\Omega^2)\ ,\label{ssol}
\eea
where $\Delta=4/N$ and
\be
a^2=\Delta -\fft{2d\tilde d} {D-2}\ .\label{avalue}
\ee

\subsection{Supersymmetry of the multi-charge $p$-brane solitons}

     Having obtained the extremal multi-charge solutions, we may now apply
the formalism developed in the previous section for determining the
fractions of supersymmetry that are preserved by them.  To do this, it is
simply a matter of substituting the appropriate charges, and the
expression (\ref{massform}) for the mass per unit length, into the general
expression (\ref{genbog}) for the \bog matrix.  Then, an elementary
calculation gives the eigenvalues of the \bog matrix, and the fraction of
supersymmetry that is preserved is equal to the fraction of the total of
32 eigenvalues that are equal to zero.  Some caution has to be exercised
in applying this formula, and we shall comment on this further as we
proceed.

     Let us first consider solutions involving just a single charge $Q$.  In
all such cases, we see from (\ref{genbog}) that the form of the
corresponding \bog matrix will be
\be
{\cal M}= m \, \oneone + Q\, \Gamma\ ,\label{onebog}
\ee
where $\Gamma$ represents the particular product of gamma matrices
associated with the field strength that carries the charge, as given in
(\ref{genbog}).  It is clear that in all cases corresponding to $p$-branes
with $p\ge0$, the associated product of gamma matrices is 
hermitean\footnote{The products of gamma matrices of
the types $\Gamma_{ijk}$ or $\Gamma_{ij}$ are exceptions, since these are
anti-hermitean.  However, these are associated with electric
$(-1)$-branes, which are instantons whose existence requires that the
timelike coordinate be Euclideanised.  There will be an extra factor of
$i$ coming from the electric charge in such cases, which restores the
hermiticity of the \bog matrix.}, and
$\Gamma^2=\oneone$. Thus we see from (\ref{onebog}) that
$({\cal M}-m)^2=Q^2$, and hence by the Cayley-Hamilton theorem the
eigenvalues $\mu$ of the \bog matrix for single-charge solutions are
\be
\mu=m\pm Q\ ,\label{oneev}
\ee
with 16 eigenvalues for each sign choice.  From (\ref{massform}), we
therefore see that the extremal one-charge $p$-branes all have \bog
eigenvalues given by 
\be
\mu=2Q\,\{ 0_{16},1_{16}\}\ ,
\ee
where the subscripts on each term indicate their degeneracies.  Thus the
single-charge extremal $p$-branes all preserve $\ft12$ of the
supersymmetry.

   Turning now to two-charge solutions, it is easiest first to consider a
particular example.  Let us take the case of a black hole in
$D=9$ carrying two electric charges.  There are in total three 2-form
field strengths in $D=9$, namely $F_2^{12}$, ${\cal F}_2^{1}$ and ${\cal
F}_2^{2}$.  From the definitions (\ref{dilatonvec}) for their associated
dilaton vectors $\vec a_{12}$, $\vec b_1$ and $\vec b_2$, it is easy to see
that either of the pairs $\{\vec a_{12},\vec b_1\}$ or $\{\vec a_{12}, \vec
b_2\}$ satisfies the condition (\ref{mmatrix}), whilst the pair $\{\vec
b_1,\vec b_2\}$ does not.  Let us take the first case, where the two
charges are carried by the field strengths $F_2^{12}$ and ${\cal F}_2^{1}$.
Denoting these charges by $Q_1$ and $Q_2$, we have from (\ref{genbog})
that the \bog matrix is
\be
{\cal M}= m\, \oneone + Q_1\, \Gamma_{0\td 1 \td 2} + Q_2\, \Gamma_{0\td
1}\ ,
\ee
where $\td 1$ and $\td 2$ denote the index values $i$ associated with the
first and the second steps of reduction from $D=11$ to $D=9$.  Thus we have
$({\cal M}-m)^2= Q_1^2 + Q_2^2 + 2Q_1\, Q_2\, \Gamma_{\td 2}$, and, after
shifting terms and squaring again, $(({\cal M}-m)^2 -Q_1^2 -Q_2^2)^2 =
4Q_1^2\, Q_2^2$.  This implies that the eigenvalues of the \bog matrix in
this case are given by
\be
\mu= m\pm Q_1\pm Q_2\ ,\label{mbog2}
\ee
where the two $\pm$ signs are independent.  It is not hard to see that in
all the two-charge $p$-branes, the expression for the eigenvalues of the
\bog matrix will be the same.  If we now use the expression
(\ref{massform}) for the mass of the extremal two-charge solution, namely
$m=Q_1+Q_2$, we see that the eigenvalues are
\be
\mu=2\, \{0,Q_1,Q_2,Q_1+Q_2\}\ .\label{twoev}
\ee
where each eigenvalue occurs with degeneracy 8.  Thus for two-charge
extremal solutions with generic values for the charges, $\ft14$ of the
supersymmetry is preserved.  If either charge is set to zero, the
situation reduces to the previously-discussed single-charge solution, and
$\ft12$ of the supersymmetry is preserved in this case.

     At this point a word of caution is appropriate.  It might seem from
the form of (\ref{twoev}) that a supersymmetry enhancement from $\ft14$ to
$\ft12$ could also be achieved by choosing $Q_2$ to be $-Q_1$.  However,
this is in fact not the case, and the reason is that in the discussion of
the \bog matrix, and its relation to supersymmetry, it was tacitly assumed
that the class of metrics that were being discussed were free of naked
singularities.  Provided this is true, then zeroes of the \bog matrix are
associated with components of unbroken supersymmetry.  However,
bearing in mind that the charges $Q_\a$ are related to the integration
constants $\lambda_\a$ appearing in the metric (\ref{gensol1}) by
$Q_\a=\ft14 \lambda_\a$, we see that choosing any of the charges $Q_\a$
here to be negative will imply that the metric functions will become
singular for some {\it positive} value of $r$, and in fact the curvature
tensor will diverge there.\footnote{It should perhaps be emphasised that
there is really nothing special about positive rather than negative
charges here.  Our statements are made with respect to a convenient set of
conventions that we have chosen, in which we pick the $p$-brane solutions
in which positive charge means positive mass.  There are another set of
solutions where negative charge means positive mass.  Rather than increase
the complexity of all discussions by having to keep track of both sets of
solutions, we have picked just the first set, and consequently there is an
asymmetry between positive and negative charges with respect to this
subset of the solutions.}  Now the horizon of the extremal
$p$-brane lies at $r=0$, and so it follows that if any of the charges
$Q_\a$ is negative, there will be naked singularities outside the horizon. 
Under such circumstances the validity of the \bog matrix discussion in the
previous section breaks down, and in particular the association between zero
eigenvalues and unbroken supersymmetry ceases to be generally valid.  A
further illustration of the breakdown of the discussion is provided by the
fact that if either of the charges is chosen to be negative, the \bog
matrix (\ref{twoev}) will also have negative eigenvalues.  This would
contradict the fact that, subject to appropriate regularity assumptions
for the metric, its eigenvalues are always non-negative.  The resolution,
of course, is that the naked singularities violate the regularity
assumptions.

     Turning now to 3-charge solutions, it is straightforward to carry out the 
analogous steps to those described above, in order to calculate the 
eigenvalues of the \bog matrix.  Again, it turns out that the expressions for 
the eigenvalues in terms of the charges $Q_1$, $Q_2$ and $Q_3$ are the same
for all cases, and after some algebra we find that they are given by
\be
\mu=m\pm Q_1\pm Q_2\pm Q_3\ ,\label{threeev} 
\ee
where the three $\pm$ signs are independent.  Applying this formula to the
extremal 3-charge solutions, for which from (\ref{massform}) we have
$m=Q_1+Q_2+Q_3$, we see that the eigenvalues of the \bog matrix are
\be
\mu=2\, \{ 0, Q_1,Q_2,Q_3,Q_{12},Q_{13},Q_{23},Q_{123}\}\ ,\label{3ch}
\ee
with each eigenvalue occurring with degeneracy 4.  Here, we have introduced 
the the notation that $Q_{i\cdots j}\equiv Q_i+\cdots +Q_j$.
Thus all generic 3-charge solutions preserve $\ft18$ of the supersymmetry.
If one or more charges are set to zero, the results reduce to those of the
previously-discussed two-charge or one-charge solutions.  Again, any
apparent enhancement of supersymmetry achieved by taking some charges to be
negative to get further zeroes in (\ref{3ch}) is ``bogus,'' for the same
reasons we discussed above.

     One might think that the discussion would proceed uneventfully to all
$N$-charge solutions for all higher values of $N$.  However, starting with
$N=4$ it turns out that the situation becomes a little more complicated.
In particular, there are two different kinds of result that can arise for
the eigenvalues of the \bog matrix for 4-charge solutions.  In the case of
2-form field strengths, only one of these possibilities can be realised, 
although in fact this possibility itself divides into two sub-categories.
We find that the eigenvalues of the \bog matrix for 4-charge 2-form solutions
are given by
\be
\mu=m\pm Q_1\pm Q_2\pm Q_3 \pm Q_4\ ,\label{four2fev}
\ee
but in this case the $\pm$ signs are not all independent, and only 
eight combinations out of the total of 16 occur in any given case.  In fact
there are exactly two possibilities for the combinations that occur; either
it is the eight cases where there are an even number of minus signs, or it is
the other eight cases where there are an odd number of minus signs.  It is the
details of the charge configurations in a given solution that determine
which of the two possibilities is realised for that solution. 
In the case of the extremal 4-charge solutions, we therefore either obtain the
eigenvalues
\be
\mu =2\, \{0,Q_{12},Q_{13},Q_{14},Q_{23},Q_{24},Q_{34},Q_{1234}\}\ ,
\label{b1}
\ee
or else, with the other set of sign choices, we get the eigenvalues
\be
\mu=2\, \{Q_1,Q_2,Q_3,Q_4,Q_{234},Q_{134},Q_{124},Q_{123}\}\ , \label{b2}
\ee
each with degeneracy 4.  Thus the generic 4-charge solutions using 2-form field
strengths again preserve $\ft18$ of the supersymmetry, in the first choice of
sign combinations.  In the second choice, the solution will preserve no
supersymmetry at all, even though it is extremal.  Both of these
possibilities can be realised for all configurations using four 2-form field
strengths.  As we mentioned above, since the charges enter the field
equations quadratically it follows that there is a bifurcation of solutions
for each of the participating charges: in one branch positive charge
contributes positively to the mass, while in the other branch negative
charge contributes positively to the mass.  In solutions with $N\le 3$
charges all $2^N$ branches have the same supersymmetry, but when $N=4$ eight
branches give the eigenvalues (\ref{b1}) whilst the other eight give
(\ref{b2}).

     The other possibility for the structure of the \bog eigenvalues for
4-charge solutions can occur only for 1-form field
strengths.  This is the case denoted by 4, as opposed to $4'$, in table 1. 
Here, we find that the  eigenvalues of the \bog matrix are again given by the
expression  (\ref{four2fev}), except that now the $\pm$ choices are all
independent.  In  this case each eigenvalue therefore occurs with degeneracy
2.  Since all the sign combinations occur here, there is no division into two
sub-categories in this case.  Extremal 4-charge solutions of this kind have
eigenvalues
\be
\mu=2\, \{0,Q_1,Q_2,Q_3,Q_4,Q_{12},Q_{13},Q_{14},Q_{23},Q_{24},Q_{34},
Q_{234},Q_{134},Q_{124},Q_{123},Q_{1234}\}
\ee
and so they preserve $\ft1{16}$ of the supersymmetry.  In this case there
is no non-supersymmetric variant.  An example of a 4-charge solution that
gives this set of \bog eigenvalues is one using the 1-form field strengths
${\cal F}_1^{12}$, ${\cal F}_1^{34}$, ${\cal F}_1^{56}$ and $F_1^{127}$,
whereas the previous eigenvalues (\ref{b1}) and (\ref{b2}) are achieved
using, for example, ${\cal F}_1^{12}$,
${\cal F}_1^{45}$, $F_1^{123}$ and $F_1^{345}$ \cite{lpmulti}.

     For 1-form solutions with 5, 6, 7 or 8 charges we find that again there
are two possible sub-categories of eigenvalue structures, one yielding two
zero eigenvalues, thus implying that $\ft1{16}$ of the supersymmetry is
preserved, and the other yielding no zero eigenvalues.  Further details can
be found in \cite{lpmulti}. 

\subsection{Non-extremal $p$-brane solitons}

     The $p$-brane solitons that we have discussed up until now have
been extremal solutions, in which the mass per unit $p$-volume takes its
lowest possible value with respect to the charges carried by the field
strengths in the solution, while still avoiding the occurrence of naked
singularities.  In this circumstance, for which a \bog bound is
saturated, the solution typically preserves some fraction
of the supersymmetry.  However, in cases where four or more independent
field strengths carry charges, we have seen that there can also exist
solutions which, although still extremal, preserve none of the supersymmetry.
In this section, we shall discuss more general $p$-brane solutions in which
the mass per unit $p$-volume is a further free parameter, independent of the
charges.  These non-extremal, or ``black,'' $p$-branes preserve no
supersymmetry.  

     We shall discuss two different kinds of generalisation away from the
previous extremal solutions.  The first of these, giving what have
been called type-2 non-extremal $p$-branes in \cite{lmp2}, involves a
modification to the ansatz (\ref{metricans2}) for the metric.
Specifically, the new ansatz becomes \cite{dlpblack}
\be
ds^2=e^{2A}\, (-e^{2f}\, dt^2 + dx^i\, dx^i) + e^{2B}\, (e^{-2f}\, dr^2+
r^2\, d\Omega^2)\ ,\label{blackmetric}
\ee
where the new function $f$, like $A$ and $B$, depends only on $r$.
The ansatz for the field strengths remains unchanged from the extremal case.
After straightforward calculations (see, for example, \cite{dlpblack} for the
details), one finds that the function $f$ has the following universal
solution:
\be
e^{2f} = 1-\fft{k}{r^{\td d}}\ ,\label{fsoln}
\ee
and that the dilaton and metric have the solutions
\bea
&&e^{\ft12 \epsilon\vp_\a -dA} = 1 + \fft{k}{r^{\td d}} \sinh^2\mu_\a
\ ,\qquad e^{2f}=1 -\fft{k}{r^{\td d}}\ ,\nonumber\\
&&ds^2 = \prod_{\a=1}^{N} \Big (1 + \fft{k}{r^{\td d}} \sinh^2\mu_\a
\Big)^{-\ft{\td d}{D-2}} ( -e^{2f} dt^2 + dx^i dx^i)\label{msbsol}\\
&&\phantom{xxxx} + \prod_{\a=1}^N \Big (1 + \fft{k}{r^{\td d}}
\sinh^2\mu_\a \Big)^{\ft{d}{D-2}} ( e^{-2f} dr^2 + r^2d\Omega^2)\ .\nonumber
\eea

     The metrics described by (\ref{msbsol}) have an outer event horizon at
$r=k^{1/\td d}$ (assuming $k$ is positive), and in general the curvature
diverges at $r=0$.  Thus they describe $p$-brane generalisations of black
holes, in which the curvature singularity is hidden behind an horizon.  In
the limit where $k$ goes to zero, the previous extremal solutions are
recovered, in which, in general, the horizon at $r=0$ coincides with a
curvature singularity.  The mass per unit volume and the charges for this
solution are given by
\be
m = k \, (\td d\sum_{\a=1}^N \sinh^2\mu_\a + \td d +1)\ ,
\qquad Q_\a= \ft12 \td d\, k\, \sinh2\mu_\a\ ,
\label{mc3} 
\ee
where we used the ADM mass formula obtained in \cite{jxlu} for the
metric (\ref{msbsol}).  Thus the mass $m$ and the $N$ charges $Q_\a$
are parameterised in terms of the $N+1$ independent constants $k$ and
$\mu_\a$. For non-negative values of $k$, the mass and charges satisfy
the bound
\be
m - \sum_{\a=1}^N Q_\a = \ft12 k\td d \sum_{\a=1}^N (e^{-2\mu_\a} -1) +
k (\td d + 1) \ge \fft{k \td d (d-1)}{d}\ge 0\ ,\label{bound3}
\ee
which coincides with the Bogomol'nyi bound. In the extremal limit
$k\longrightarrow 0$ it is saturated, and the solutions become
supersymmetric.  When the parameters are chosen such that the mass exceeds
the bound (\ref{bound3}), the \bog matrix has only positive eigenvalues, as
can be seen explicitly from our formulae (\ref{oneev}), (\ref{mbog2}),
(\ref{threeev}) and (\ref{four2fev}) in the cases of $N=1,2,3$ and 4 charges.

     As in the discussion of the extremal multi-charge solutions in section
4.1, we may again consider the special case where all $N$ charges are set
equal.  This gives a solution which also solves the reduced single scalar,
single field strength system (\ref{slag}), with
\vfill\eject
 
\bea
ds^2&=& \Big( 1+ \fft{k}{r^{\td d}} \sinh^2\mu\Big)^{-\ft{4 
\td d}{\Delta(D-2)}} (-e^{2f}dt^2 + dx^idx^i) \nonumber\\
&&+\Big( 1+ \fft{k}{r^{\td d}} \sinh^2\mu\Big)^{\ft{4
d}{\Delta(D-2)}} (e^{-2f} dr^2 + r^2 d\Omega^2)\ ,\nonumber\\
e^{\ft{\epsilon\Delta}{2a} \phi} &=& 1 + \fft{k}{r^{\td d}} \sinh^2\mu
\ ,\qquad e^{2f} = 1 - \fft{k}{r^{\td d}}\ ,\label{sssol}
\eea
where $\Delta=4/N$ and $a$ is given by (\ref{avalue}).  The two free
parameters $k$ and $\mu$ are related to the charge $Q$ and the mass
per unit $p$-volume $m$ by
\be
Q= \fft{\td d k}{\sqrt\Delta} \sinh2\mu\ ,\qquad
m= k \Big(\fft{4\td d}{\Delta} \sinh^2\mu + \td d+1\Big)
\ .\label{mc1}
\ee

     There is also another kind of generalisation away from the extremal
$p$-brane solitons \cite{lpxtoda}, giving rise to what have been
called type-1 non-extremal $p$-branes in \cite{lmp2}.  In this case
the metric ansatz (\ref{metricans2}) remains unchanged from its
extremal form.  The change from the procedure that gives the extremal
$p$-branes comes as a result of not imposing any further restriction
on the various functions in the ans\"atze for the metric, dilaton and
field strengths, but instead constructing the most general solution of
the equations of motion.  In particular, the relation $dA + \td d B=0$
for the metric functions in (\ref{metricans2}) is no longer imposed.
Accordingly, we begin by defining
\bea
&&X = dA + \td d B \ ,\qquad Y = A + \fft{\epsilon \td d}{D-2} 
\sum_{\a,\beta} (M^{-1})_{\a\b} \varphi_\a\ ,\nonumber\\
&&\Phi_\a = \epsilon \varphi_\a - 2d A\ .\label{multivar}
\eea
It is advantageous to introduce a new radial variable $\rho=r^{-\td d}$, in
terms of which the equation of motion for $X$ and $Y$ turn out to be 
\be
\fft{d^2 X}{d\rho^2} +\Big(\fft{d X}{d\rho}\Big)^2 -\fft1{\rho}\,
\fft{dX}{d\rho} =0\ ,\qquad \fft{d^2 Y}{d\rho^2} +
\fft{dX}{d\rho}\, \fft{dY}{d\rho} =0\ ,
\ee
giving the solutions $e^X=1-k^2\, \rho^2$ and $Y=-(\mu/k)\,
{\rm arctanh}(k\rho)$. The further change of radial variable to $\xi$ 
defined by
$k\, \rho =
\tanh(k\, \xi)$ reduces the remaining equations of motion to \cite{lpxtoda}
\bea
&&\ddot\Phi_\a = -\fft{32 Q_\a^2}{\td d^2} 
e^{-\Phi_\a}\ , \label{multiliou}\\ 
&&\sum_{\a=1}^{N} \Big(\fft{32 Q_\a^2}{\td d}
e^{-\Phi_\a}- \ft12 \td d \dot
\Phi_\a^2 \Big) = -16(\td d +1) k^2 + 2d \Big(2(D-2) - 
d \td d N\Big)\mu^2\ ,\label{multicon} 
\eea
provided that the dilaton vectors for the participating field strengths
satisfy the relations (\ref{mmatrix}).  Note that a dot denotes a derivative
with respect to the redefined radial variable $\xi$. In terms of $\xi$, the
solutions for $X$ and $Y$ are
\be
e^{-\ft12 X}= \cosh k\xi \ ,\qquad Y=-\mu\, \xi\ .\label{xysol}
\ee

     The equations (\ref{multiliou}) are $N$
independent Liouville equations for the functions $\Phi_\a$, subject to the
single first-integral constraint (\ref{multicon}).  This can be
re-expressed in terms of the Hamiltonian
\be
H\equiv \sum_{\a=1}^N\Big(\ft12 p_\a^2 - \fft{32Q_\a^2}{\td d^2}\,
e^{-\Phi_\a}\Big) = \fft{16(\td d+1)}{\td d} k^2 - \fft{2d}{\td d}\,
\Big(2(D-2) - d \td d N\Big) \mu^2\ ,
\ee
where $p_\a$ is the momentum conjugate to $\Phi_\a$.  Hamilton's equations
then give (\ref{multiliou}).

   The solutions of the Liouville equations (\ref{multiliou}) for $\Phi_\a$
imply that 
\be
e^{\ft12\epsilon \varphi_\a - d A} = \fft{4 Q_\a}{\td d \beta_\a}
\sinh(\beta_\a \xi + \gamma_\a)\ ,
\ee
where $\beta_\a$ and $\gamma_\a$ are constants, 
while (\ref{multicon}) gives the constraint $\td d\sum_\a \beta_\a^2 = 
-8(\td d+1) k^2 + d(2(D-2) -  d \td d N)\mu^2$.  The solutions for the
functions $A$ and $B$ in the metric (\ref{metricans2}) can be written as
\cite{lpxtoda}
\bea
e^{-2(D-2) A/\td d} &=& e^{(2(D-2)-d\td d N)\mu\xi/\td d}
\prod_{\a=1}^{N} \Big (\fft{4Q_\a}{\td d \beta_\a} \sinh(\beta_\a\xi + 
\gamma_\a)\Big)\ ,\\
e^{2(D-2) B/d} &=& (\cosh(k\xi))^{-(D-2)/(d\td d)}
e^{(2(D-2)-d\td d N)\mu\xi/\td d}
\prod_{\a=1}^{N} \Big (\fft{4Q_\a}{\td d \beta_\a} \sinh(\beta_\a\xi + 
\gamma_\a)\Big)\ .\nonumber
\eea
The solutions have an outer
event horizon at $r^{\td d} = k$  ({\it i.e.}\ at $\xi=\infty$). The mass
per unit $p$-volume is given by
\be
m=2(D-2)\mu -d\td d\, N\, \mu +\sum_\a Q_\a\, \cosh\gamma_\a\ ,
\ee
where the integration constants $\gamma_\a$ are chosen such that
$4 Q_\a\, \sinh\gamma_\a=\td d\, \beta_\a$, so that the metric approaches
the standard Minkowski metric at infinity.

     It is clear that when $p$ is greater than zero, the type-1
non-extremal $p$-brane solutions given by this construction are quite
different from the type-2 solutions described by (\ref{msbsol}). In
particular, in the type-2 solutions the spacetime metric on the $p$-brane 
world-volume is no longer Poincar\'e invariant, owing to the extra $e^{2f}$
factor in front of $dt^2$ in (\ref{blackmetric}).   By contrast, the metric
for the type-1 solutions has the same fully Poincar\'e invariant form
(\ref{metricans2}) as for the extremal $p$-branes.  However, in the special
case that $p=0$, it is clear that the two metric ans\"atze
(\ref{blackmetric}) and (\ref{metricans2}) are simply related by a
coordinate transformation of the radial variable $r$, and so in the special
case of black holes, the type-1 non-extremal solutions encompass the type-2
ones.

\subsection{Dyonic $p$-brane solutions}

     If the spacetime dimension $D$ is even, it is possible that a
field strength of degree $n=\ft12 D$ can carry both electric and magnetic
charge at the same time.  In such cases, the possibility of having
dyonic $p$-brane solutions arises \footnote{There is another kind of
solution that is sometimes called dyonic, in which two or more field
strengths carry charges, some of them electric and the others magnetic. 
These are not really intrinsically dyonic, since they can be rendered purely
electric or purely magnetic by dualisations.}.  Since we are considering the
toroidal compactifications of M-theory, the dimensions in which this might
occur are $D=8$, 6 and 4.  We shall postpone the discussion of $D=8$ until
the end of this section, and consider $D=6$ and $D=4$ first.  The dyonic
solutions arise in the case where just one of the field strengths, of
degree $n=3$ in $D=6$ or degree $n=2$ in $D=4$, is non-zero, and thus the
configurations satisfy the equations of motion from the reduced
single-scalar system (\ref{slag}).  Let us begin by considering the
general equations of motion for this system, with the metric ansatz
(\ref{metricans2}) and the two field-strength ans\"atze (\ref{fansatz})
imposed simultaneously, so that $F$ carries both electric and magnetic
charge. Introducing the redefined radial
coordinate $\xi$, one finds that $X$ and $Y$ have the
same solutions as in section 4.3, and the remaining equations of motion can
be cast into the form \cite{lpsol}
\bea
&&\ddot q_1 = e^{\a q_1 + (1-\a) q_2}\ ,\qquad
\ddot q_2 = e^{(1-\a)q_1 + \a q_2}\ ,\label{toeq}\\
&&H \equiv \fft{\a}{2(2\a-1)}\, (p_1^2+ p_2^2) + \fft{\a-1}{2\a-1}\,
p_1 p_2\nonumber
 - e^{\a q_1 - (1-\a) q_2} - e^{(1-\a) q_1 + \a q_2}
=8nk^2\ ,\label{todahamil}
\eea
where 
\be
A=\fft1{4(n-1)}\Big(q_1 + q_2 -2\log\fft{\lambda_1\lambda_2}{n-1}\Big)\ ,
\qquad\phi = \fft{a}{2(n-1)} (q_1 -q_2) + 
\fft{1}{a}\log\fft{\lambda_2}{\lambda_1}\ ,\label{redef1}
\ee
the constant $\a$ is related to $a$ by
\be
\a = \ft12 + \fft{a^2}{2(n-1)} = \fft{\Delta}{2(n-1)}\ ,
\ee
and $H=H(p_1,p_2,q_1,q_2)$ is the Hamiltonian.  Thus Hamilton's equations
$q_i' = \del H /\del p_i$ imply that
\be
p_1 = \a \dot q_1 + (1-\a) \dot q_2\ ,\qquad
p_2 = (1-\a) \dot q_1 + \a \dot q_2\ ,
\ee
while $\dot p_i = -\del H/\del q_i$ gives precisely the equations of motion
(\ref{toeq}). 

     As far as we know, the general solution to the equations (\ref{toeq})
cannot be given in closed form except for two special values of $\a$, namely 
\bea
\a=1:&& \ddot q_1 = e^{q_1} \ ,\qquad\quad\ \ 
\ddot q_2 = e^{q_2}\ ,\nonumber\\
\a = 2:&& \ddot q_1 = e^{2q_1 - q_2} \ ,\qquad \ddot q_2 = e^{2q_2 -q_1}\ .
\label{lioutod}
\eea
The first case gives two independent Liouville equations, while the second
gives the $SL(3,R)$ Toda equations.  They correspond to values of $\Delta$
that are allowed in the maximal supergravities in $D=6$ and $D=4$
respectively, namely $\Delta=4$ in each case.  Since the allowed values
take the form $\Delta=4/N$, we see that indeed, as stated above, these two
solvable cases involve just one field strength.

     The dyonic string solution in $D=6$, where the equations separate as
two Liouville equations, is easily found to be \cite{lpxtoda}
\bea
e^{-\phi/\sqrt2 - 2 A} &=& \fft{2Q_m}{\beta_1}
\sinh(\beta_1 \xi + \gamma_1) \ ,\nonumber\\
e^{\phi/\sqrt2 - 2 A} &=& \fft{2Q_e}{\beta_2}
\sinh(\beta_1 \xi + \gamma_1) \ ,
\eea
with the constraint $\beta_1^2 + \beta_2^2 = 4 nk^2$, where $Q_e$ and
$Q_m$ are the electric and magnetic charges of the string.  The solution has
an outer event horizon at $\rho = 1/k$ ({\it i.e.}\ at $\xi =\infty$), and
the mass per unit length is
\be
m=\sqrt{Q_e^2 +\ft32 k^2} + \sqrt{Q_m^2 +\ft32 k^2}\ ,
\ee
where we have chosen $2Q_m\, \sinh\gamma_1 = \beta_1$ and $2Q_e\,
\sinh\gamma_2 = \beta_2$ so that the solution approaches the
standard Minkowski spacetime at infinity, and the dilaton vanishes there. 
The usual extremal dyonic string \cite{dfkr} is recovered in the limit 
when $k$
goes to zero.  The eigenvalues of the \bog matrix in the case of this dyonic
string are
\be
\mu= m\pm Q_e\pm Q_m\ ,\label{strbog}
\ee
where the $\pm$ signs are independent, and thus in the extremal limit where
\be
m=Q_e+Q_m\label{d6dyonmass}
\ee
we have $\mu=2\{0,Q_e,Q_m,Q_e+Q_m\}$ and the solution preserves
$\ft14$ of the supersymmetry.  As usual, the occurrence of 8
further zero eigenvalues when $Q_m=-Q_e$ does not imply any enhancement of
the supersymmetry since the solution then has naked singularities and the
\bog analysis becomes invalid. 

     Before moving on to the dyonic Toda black hole in $D=4$, we should
remark that in addition to the type-1 non-extremal dyonic string obtained
above, there is also a more standard type-2 non-extremal solution, where
the metric ansatz has the form (\ref{blackmetric}), and $B=-A$.  The
solutions for $\phi$ and $A$ are given by
\cite{dlpblack}
\be
e^{-\phi/\sqrt2 -2 A} = 1 + \fft{k}{r^2} \sinh^2\mu_1\ ,\qquad
e^{\phi/\sqrt2 -2 A} = 1 + \fft{k}{r^2} \sinh^2\mu_2\ ,
\label{dyon}
\ee
with $f$ as usual given by (\ref{fsoln}).  The mass per unit length and the
charges are given in terms of $k$ and $\mu_i$ by
\be
m = k (2\sinh^2\mu_1 + 2\sinh^2\mu_2 + 1)\ ,\qquad
Q_m =k\,\sinh 2\mu_1 \ ,\qquad Q_e =k\,\sinh 2\mu_2 .\label{mc2}
\ee
For non-negative values of $k$, the mass and the charges satisfy the 
bound
\be
m-Q_e-Q_m = k  +k\,  e^{-2\mu_1} +k\,  e^{-2\mu_2} \ge 0\ .
\label{bound2}
\ee
The bound is saturated in the extremal limit $k\longrightarrow 0$.

     Turning now to the dyonic black hole in $D=4$, for which the
equations of motion reduce to the $\a=2$ case in (\ref{lioutod}), one finds
from the general solution of the Toda equation that the solutions for
$\phi$ and $A$ are given in terms of four arbitrary constants $c_1$,
$c_2$, $\mu_1$ and $\mu_2$ by \cite{lpxtoda}
\bea
Q_m^{-4/3}Q_e^{-2/3}\, e^{- \phi/\sqrt3 - 2 A}
&=& \fft{16 c_1 e^{\mu_1 \xi}}{\nu_1(\nu_1 -\nu_2)} -
\fft{16 c_2e^{\mu_2 \xi}}{\nu_2(\nu_1 -\nu_2)} +
\fft{16 e^{-(\mu_1+\mu_2) \xi}}{c_1 c_2\nu_1\nu_2}\ ,
\nonumber\\
Q_e^{-4/3}Q_m^{-2/3} \,e^{\phi/\sqrt3  - 2 A}
&=& \fft{16 e^{-\mu_1 \xi}}{c_1 \nu_1(\nu_1 -\nu_2)} -
\fft{16 e^{-\mu_2 \xi}}{c_2 \nu_2(\nu_1 -\nu_2)} -
\fft{16 c_1c_2e^{(\mu_1 +\mu_2) \xi}}{\nu_1\nu_2}\ ,\label{a3case}
\eea
where $\nu_1 = 2\mu_1 + \mu_2$ and $\nu_2 = 2\mu_2 + \mu_1$, together with 
the constraint $H=\mu_1^2 + \mu^2_2 + \mu_1 \mu_2 = 16k^2$.  

     The extremal limit of the dyonic black hole can be found by taking $k$
to zero appropriately in the above solution.  An easier way of obtaining
the extremal solution is by directly re-solving the Toda equations subject
to the Hamiltonian constraint (\ref{todahamil}) with $k=0$.  The required
solution is obtained by making the ansatz that $e^{-q_2}=e^{-q_1} +$const.
With this ansatz, it is easy to verify that $(e^{-q_1})''=(e^{-q_2})''=1$,
where here a prime denotes a derivative with respect to the redefined
radial variable $\rho=1/r$.  Thus, in terms of the original variables
$\phi$ and $A$, the solution takes the form
\bea
e^{\phi/\sqrt3-2A} &\equiv& T_m= 1 + 4 Q_m^{2/3}\, (Q_e^{2/3} + 
Q_m^{2/3}) \,
\fft1{r} + 8 Q_e^{2/3}\, Q_m^{4/3}\, \fft1{r^2}\ ,\nn\\
e^{-\phi/\sqrt3-2A} &\equiv& T_e =  1 + 4 Q_e^{2/3}\, (Q_e^{2/3} + 
Q_m^{2/3}) \,
\fft1{r} + 8 Q_m^{2/3}\, Q_e^{4/3}\, \fft1{r^2}\ ,\label{dyonsol}
\eea
where we have chosen certain integration constants so that $\phi$ and $A$
approach zero as $r$ tends to infinity.  The metric of the extremal dyonic
black hole is given by
\be
ds^2= -(T_e\, T_m)^{-1/2}\, dt^2 + (T_e\, T_m)^{1/2}\, (dr^2 + r^2\,
d\Omega^2)\ .\label{d4dyonmetric} 
\ee
An interesting feature of this solution is that the mass is given in terms of
the electric and magnetic charges by the curious formula \cite{gk}
\be
m=\Big( Q_e^{2/3} + Q_m^{2/3}\Big)^{3/2}\ .\label{dyonmass}
\ee
Since the \bog matrix in this case has eigenvalues $\mu=m\pm Q_e\pm Q_m$,
it follows that even in this extremal limit, the solution has no
supersymmetry (unless $Q_e=0$ or $Q_m=0$):  It is easily seen that the
eigenvalues are strictly positive unless one of the charges vanishes.

     The final example of a dyonic solution in toroidally-compactified
M-theory arises in the eight-dimensional theory.  Here, there the 4-form
field strength $F_4$ can carry both electric and magnetic charge, giving
rise to a dyonic membrane solution \cite{ilpt}.  This solution is rather
different from the previous ones we have discussed, in that it not only
involves the 4-form field strength and a dilatonic scalar, but also the
0-form potential $A_0^{(123)}$.  This is clear from the form of the cubic
$FFA$ terms in $D=8$, given in (\ref{ffaterms}), which imply that
$A_0^{(123)}$ will have $F_{\sst{MNPQ}}\, F_{\sst{RSTU}}\,
\epsilon^{\sst{MNPQRSTU}}$ as a source on the right-hand side of its field
equation.  When $F_4$ carries both electric and magnetic charge, this
source will be non-zero.  In fact, the dyonic membrane solution can be
obtained by performing a duality rotation on a simple purely electric or
purely magnetic membrane solution of the standard kind.  In this respect,
the situation is quite different from that for the $D=6$ dyonic string or
the $D=4$ dyonic black hole described above, where the dyonic solutions are
not simply related to previously-known purely electric or purely magnetic
ones by duality rotations.  The forms of the metric, dilaton and axion in
the dyonic membrane solution are given by \cite{ilpt}:
\bea
ds^2 &=& H^{-\ft12} (-dt^2 + dx_1^2 + dx_2^2) +
         H^{\ft12} dy^m dy^m\ ,\nonumber\\
F_4&=&\fft12 (*dH)\,\cos \delta + \ft12 dH^{-1} \wedge dt\wedge
dx_1 \wedge dz\, \sin\delta \ ,\label{d8dyon}\\
A_0 + \im e^{2\sigma} &=& \fft{ (1-H) \sin 2\delta + 
2 \im H^{\ft12}}{ 2(\sin^2\delta +H \cos^2\delta)}\ ,\nonumber
\eea
where $\sigma = \phi_1/2 + 3\phi_2/\sqrt7 + 6\phi_3/\sqrt{21}$, and
the two orthogonal combinations of $\phi_1$, $\phi_2$ and $\phi_3$ are
zero.  The angle $\delta$ parameterises the duality rotation.  Since the
U-duality symmetry commutes with supersymmetry, the solution preserves the
same fraction $\ft12$ of the supersymmetry as in the  pure electric
and pure magnetic cases.  This can also be seen from the Bogomol'nyi bound,
which is saturated by the solution (\ref{d8dyon}) since the mass per unit
2-volume is given by
\be
m= \sqrt{Q_e^2 + Q_m^2}\ ,\label{d8dyonmass}
\ee
where $Q_e = Q\sin\delta$ and $Q_m = Q \cos\delta$, and $Q$ is the purely
magnetic charge before the U-duality rotation.

\subsection{$SL(N+1,R)$ Toda solitons}

     Before leaving the subject of $p$-brane soliton solutions, we shall
consider one further class of solutions that arises in
toroidally-compactified M-theory.  As we have indicated, the conditions
that govern whether a particular set of field strengths can be active in a
multiple-charge solution are quite stringent.  For example, we have seen
that if one restricts attention to single-scalar solutions, then in general
these can only occur if the constant $\Delta$ defined in (\ref{avalue}) is
of the form $\Delta=4/N$, where $N$ is an integer.\footnote{In \cite{lpsol},
various solutions with other values of $\Delta$, such as $\Delta=3$ were
described.  These would be perfectly valid solutions if the field strengths
appearing in (\ref{dgenlag}) were all simply the exterior derivatives of
potentials.  However, the Chern-Simons modifications that arise from the
dimensional reduction procedure complicate matters considerably, and in
particular, in general they rule out such other values of $\Delta$ in the
toroidally-compactified supergravity theories.}  There is, however, one
additional class of exceptional cases where solutions in the
toroidally-compactified supergravity theories can arise, which, when the
charges are set equal, give values of
$\Delta$ other than $4/N$.  These occur for solutions using 1-form field
strengths, which will be either $(D-3)$-branes if the field strengths carry
magnetic charges, or $(-1)$-branes (\ie instantons, for which the time
coordinate must be Euclideanised) if they carry electric charges.

     The solutions of the kind we are discussing here arise if the dilaton
vectors $\vec c_\a$ for a set of $N$ 1-form field strengths satisfy the
dot-product relations \cite{lpsol}
\be
M_{\a\beta}\equiv \vec c_\a \cdot \vec c_\beta = 4\delta_{\a\b}
-2\delta_{\a,\beta+1} - 2\delta_{\a,\beta-1}\ ,\label{cart}
\ee
which is in fact twice the Cartan matrix for $SL(N+1,R)$.  It is
straightforward to verify from the expressions given in (\ref{dilatonvec})
that there are indeed sets of 1-form field strengths whose dilaton vectors
satisfy (\ref{cart}), namely those of the form ${\cal F}_1^{i,i+1}$.  The
remarkable thing is that, as can be verified from (\ref{A.6}), these
particular field strengths have no Chern-Simons modifications.  Thus we may
consider a set of $N$ 1-form field strengths ${\cal F}_\a \equiv {\cal
F}_1^{\a,\a+1}$ whose dilaton vectors satisfy (\ref{cart}) and which are
given simply by ${\cal F}_\a=d\chi_\a$.  In $D$ dimensions, we can clearly
have up to $N_{\rm max}=10-D$ such 1-forms.  The Lagrangian (\ref{dgenlag})
can then be consistently truncated to \cite{lpsln}
\be
{\cal L} = e\, R -\ft12 e\, \sum_{\a,\b=1}^N (M^{-1})_{\a\b}\, \del_{\sst M}
\varphi_\a\, \del^{\sst M}\varphi_\b -\ft12 e\, \sum_{\a=1}^N
e^{-\varphi_\a}\, (\del\chi_\a)^2\ .\label{ttlag}
\ee

 We 
proceed by making the standard metric and magnetic field strength ans\"atze,
which in this case is
\bea
ds^2 &=& \eta_{\mu\nu} dx^\mu dx^\nu + e^{2B(r)}\, (dr^2 + r^2 d\theta^2 )
\ ,\nonumber\\
\chi_\a &=& 4Q_\a \, \theta\ .\label{ans}
\eea
Substituting into the equations of motion following from (\ref{ttlag}), we
obtain
\bea
&&\varphi_\a'' = -8\sum_\b M_{\a\b}\, Q_\b^2\, e^{-\varphi_\b}\ ,\qquad
B=\sum_{\a,\b} (M^{-1})_{\a\b}\, \varphi_\a\ ,\label{eqs0}\\ 
&&\sum_{\a,\b}(M^{-1})_{\a\b}\, \varphi_\a'\, \varphi_\b' = 16
\sum_\a Q_\a^2\, e^{-\varphi_\a} \ ,\label{eqs1}
\eea
where a prime denotes a derivative with respect to $\rho=\log r$.  
Making the redefinition  $\Phi_\a =
-2 \sum_\b (M^{-1})_{\a\b}\, \varphi_\b$, these equations become
\bea
&& \Phi_\a'' = 16 Q_\a^2\, \exp(\ft12 \sum_\b M_{\a\b}\, \Phi_\b)\ ,
\qquad B=-\ft12 \sum_\a \Phi_\a\ ,\nonumber\\
&& \sum_{\a,\b} M_{\a\b} \Phi_\a'\, \Phi_\b' = 64 \sum_\a Q_\a^2 \, 
\exp(\ft12 \sum_\b M_{\a\b}\, \Phi_\b)\ .
\eea
The further redefinition $\Phi_\a=q_\a -4 \sum_\b (M^{-1})_{\a\b} 
\log(4Q_\b)$ removes the charges from the equations, giving \cite{lpsln}
\bea
q_1''&=& e^{2q_1 -q_2}\ ,\nonumber\\
q_2''&=& e^{-q_1+2q_2-q_3}\ ,\nonumber\\
q_3''&=& e^{-q_2 + 2q_3 -q_4}\ ,\label{suntoda}\\
&&\cdots \nonumber\\
q_{\sst N}'' &=& e^{-q_{\sst{N}-1} + 2q_{\sst N}}\ .\nonumber
\eea
These are precisely the $SL(N+1,R)$ Toda equations.  The solution is subject
to the further constraint (\ref{eqs1}), which, in terms of the $q_\a$, 
becomes the constraint that the Hamiltonian
\be
{\cal H} = 4\sum_{\a,\b} (M^{-1})_{\a\b}\, p_\a\, p_\b -
           \sum_\a \exp(\ft12 \sum_\b M_{\a\b}\, q_\b) \label{hamil}
\ee
for the Toda system (\ref{suntoda}) vanishes.   

     The general solution to the $SL(N+1,R)$ Toda equations is presented
in an elegant form in \cite{ls,a}:
\be
e^{-q_\a}= \sum_{k_1<\cdots < k_\a} f_{k_1}\cdots f_{k_\a} \,
\Delta^2(k_1,\ldots,k_\a)\, e^{(\mu_{k_1}+\cdots \mu_{k_\a})\rho}\ ,
\label{gensol}
\ee
where $\Delta^2(k_1,\ldots,k_\a)=\prod_{k_i <k_j} (\mu_{k_i}- 
\mu_{k_j})^2$ is the Vandermonde determinant, and $f_k$ and $\mu_k$ are
arbitrary constants satisfying
\be
\prod_{k=1}^{N+1} f_k = - \Delta^{-2}(1,2,\ldots,N+1)\ ,\qquad
\sum_{k=1}^{N+1} \mu_k =0\ .
\ee
The Hamiltonian, which is conserved, takes the value ${\cal H}=\ft12 
\sum_{k=1}^{N+1} \mu_k^2$.

     The solution (\ref{gensol}) in general involves exponential  
functions of $\rho$.  Furthermore, the vanishing of the Hamiltonian implies
that the parameters $\mu_k$, and hence the solutions, will in general be
complex.  However, there exists a limit, under which all the $\mu_k$ 
constants vanish, which achieves a vanishing Hamiltonian and real solutions
that are finite polynomials in $\rho$.  Since we are constructing 
$(D-3)$-branes in $D\ge 3$, it follows that we are interested in obtaining
solutions to the $SL(N+1,R)$ Toda equations for $N\le 7$.  When $N=1$, the
Toda system reduces to the Liouville equation, giving rise to the usual
single field strength solution that preserves $1/2$ the supersymmetry, namely
\be
e^{-q_1}=1 + 4 Q \,\rho \ . \label{liousol}
\ee
Note that since there is only a single independent $\mu$ 
parameter when $N=1$, which has to be zero by the Hamiltonian constraint, 
(\ref{liousol}) is in fact the only solution in this case.
   
     For $N=2$, we find that the polynomial solution to the $SL(3,R)$ Toda 
equations (\ref{suntoda}) is
\bea
e^{-q_1} &=& a_0 + a_1\, \rho + \ft12 \, \rho^2\ ,\nonumber\\
e^{-q_2} &=& a_1^2 - a_0 + a_1\, \rho + \ft12 \, \rho^2\ ,
\eea
where $a_0$ and $a_1$ are constants that are related to the charge parameters
$Q_1$ and $Q_2$, on using the boundary condition that the dilatonic scalars,
and hence $\Phi_\a$, vanish ``asymptotically'' ({\it i.e.\ }at $\rho=0$).  
Thus we have 
\be
a_0= \ft{1}{16} Q_1^{-4/3}\, Q_2^{-2/3}\ , 
\qquad a_1=\ft14  Q_1^{-2/3}\,  Q_2^{-2/3}\,
( Q_1^{2/3}+ Q_2^{2/3})^{1/2} \ ,
\ee
which implies that the metric is
\be
ds^2 = \eta_{\mu\nu}dx^\mu dx^\nu + T_1 T_2 (dr^2 + r^2 d\theta^2)
\ ,\label{su3metric}
\ee
where 
\bea
T_1 &=& 1+ 4 Q_1^{2/3}\, ( Q_1^{2/3}+ Q_2^{2/3})^{1/2}\, \rho
+ 8 Q_1^{4/3}\, Q_2^{2/3}\, \rho^2\ ,\nonumber\\
T_2 &=&1+ 4 Q_2^{2/3}\, ( Q_1^{2/3}+ Q_2^{2/3})^{1/2}\, \rho
+ 8 Q_2^{4/3}\, Q_1^{2/3}\, \rho^2\ ,\label{ts}
\eea
and $\rho=\log r$.
The mass per unit $(D-3)$-volume is given by
\be
m= ( Q_1^{2/3}+ Q_2^{2/3})^{3/2}\ .\label{su3mass}
\ee
This is the same rather unusual looking mass formula that arises in
the 
$a=\sqrt3$ four-dimensional dyonic black hole \cite{gk}, which we
discussed in the previous section, and which was also associated with a
solution of the $SL(3,R)$ Toda equations.  For non-vanishing Hamiltonian the
black hole is non-extremal, becoming extremal when the Hamiltonian
vanishes.  The mass formula (\ref{su3mass}) implies that the solution
describes a system with negative binding energy, since the total mass of the
widely-separated constituents is given by
$m_\infty=Q_1+Q_2$, which is  smaller than $m$.  The \bog
matrix in this case is ${\cal M}=m\oneone + Q_1 \Gamma_{\hat1\hat2 1 2} +
Q_2 \Gamma_{\hat1\hat2 2 3}$, and therefore its eigenvalues are
\be
\mu= m \pm \sqrt{Q_1^2 + Q_2^2}\ .
\ee
It follows from (\ref{su3mass}) that the $\mu$ is strictly positive, and
hence the \bog bound is exceeded and there is no supersymmetry, unless either
$Q_1$ or $Q_2$ vanishes.

    For $N=3$, we find the following polynomial solution of the $SL(4,R)$
Toda equations:
\bea
e^{-q_1}&=& a_0 + a_1\, \rho + a_2\, \rho^2 + \ft16 \rho^3\ ,\nonumber\\
e^{-q_2}&=& a_1^2 -2 a_0 a_2 + (2 a_1 a_2 -a_0)\, \rho + 
2 a_2^2 \, \rho^2 + \ft23 a_2\,  \rho^3\  + \ft1{12} \rho^4\ ,\label{su4sol}\\
e^{-q_3}&=& a_0-4 a_1 a_2 + 8a_2^3  +(4 a_2^2 - a_1)\, \rho + a_2\, \rho^2 
+ \ft16 \rho^3\ ,\nonumber
\eea
where the constants $a_0$, $a_1$ and $a_2$ are determined in terms of the 
charges $Q_1$, $Q_2$ and $Q_3$ by the requirement that the dilatonic scalars
vanish at $\rho=0$.  This implies that
\be
e^{q_\a(0)} = \prod_\b (4 Q_\b)^{4 (M^{-1})_{\a\b}}\ ,\label{sunconst}
\ee
and hence
\bea
&&a_0= \fft1{64} Q_1^{-3/2} \, Q_2^{-1}\, Q_3^{-1/2}\ ,\qquad
a_1^2 -2 a_0 a_2= \fft1{256} Q_1^{-1}\, Q_2^{-2}\, Q_3^{-1}\ ,\nonumber\\
&&a_0-4a_1 a_2 +8a_2^3 =\fft1{64} Q_1^{-1/2} \, Q_2^{-1}\, Q_3^{-3/2}\ .
\label{3charge}
\eea
The metric is given by (\ref{ans}), with
\be
e^{2B}=\prod_\a e^{q_\a(0)-q_\a} \ ,\label{sunmetric}
\ee
and hence
\bea
m=\fft{a_1}{4 a_0} + \fft{2a_1 a_2-a_0}{4(a_1^2-2a_0 a_2)} + 
\fft{4a_2^2 -a_1}{4(a_0-4 a_1 a_2 + 8a_2^3)}\ .\label{3mass}
\eea
Thus we find \cite{lpsln} that the mass is given in terms of the charges by
the positive  root of the sextic
\vfill\eject
\bea
&&m^6- (3Q_1^2 + 2 Q_1 Q_3 + 3 Q_3^2 + 3Q_2^2)m^4 -36 \sqrt{Q_1 Q_3}Q_2
(Q_1 +Q_3) m^3 \nonumber \\
&&+\Big[(Q_1+Q_3)^2 (3Q_1^2 -2 Q_1 Q_3 + 3Q_3^2) - Q_2^2 (21 Q_1^2 + 122 
Q_1 Q_3 + 21 Q_3^2) + 3 Q_2^4 \Big] m^2 \nonumber \\
&&+ 4\sqrt{Q_1 Q_3} Q_2 (Q_1+Q_3) 
(9 Q_1^2 -14 Q_1 Q_3 + 9Q_3^2 -18Q_2^2) m\label{dgsbrkfst}\\
&&-(Q_1-Q_3)^2(Q_1+Q_3)^4 -Q_2^2(3Q_1^4-68 Q_1^3 Q_3 +114 Q_1^2 Q_3^2 -
68 Q_1 Q_3^2 + 3Q_3^4)\nonumber\\
&& -Q_2^4(3Q_1^2 + 38 Q_1 Q_3 + 3Q_3^2) -Q_2^6 =0\ .\nonumber
\eea
There seems to be  no way to give an explicit closed-form expression for 
the mass in terms of
the charges.  The \bog matrix ${\cal M}= m\oneone + Q_1 \Gamma_{\hat1\hat2 12}
+ Q_2\Gamma_{\hat1\hat2 2 3} + Q_3 \Gamma_{\hat1 \hat2 3 4}$ has eigenvalues
\be
\mu=m\pm \sqrt{(Q_1\pm Q_3)^2 + Q_2^2} \ ,
\ee
where the two $\pm$ signs are independent.  For generic values of the charges,
$\mu>0$ and the solution has no supersymmetry.  If $Q_2=0$, the solution
reduces to the two-charge supersymmetric solution, preserving $\ft14$ of the
supersymmetry.  In this case, the $SL(4,R)$ Toda equations reduce to two
decoupled Liouville equations.

     For higher values of $N$, the explicit forms of the polynomial solutions
to the $SL(N+1,R)$ Toda equations become increasingly complicated
\cite{lpsln}.  The  structure of these polynomials can be summarised as
follows.  For each
$N$,  we find that $e^{-q_\a}$ are polynomials in $\rho$ of degree
$n_\a=\a(N+1-\a)$, {\it i.e.}
\be
\fft{d^{n_\a+1}}{d\rho^{n_\a+1}} e^{-q_\a}= 0\ .
\ee
After substituting these into the $SL(N+1,R)$ Toda equations (\ref{suntoda}),
we find that there are $N$ independent parameters, which can be related to
the $N$ charges $Q_\a$ by equation (\ref{sunconst}).  The metric is given by 
(\ref{ans}) with $e^{2B}$ again given by (\ref{sunmetric}).  The mass is given
in terms of the charges by an $N!$'th-order polynomial equation.  Although it
appears not to be possible to give closed-form expressions for the mass in
terms of the charges for $N\ge3$, we expect nevertheless that it is less than
the sum of the charges, indicating again that they are bound states with
negative binding energies.  One can see this explicitly in the special case
where the charges have the fixed ratio given by
\be
Q_\a = a Q \Big(\sum_\b (M^{-1})_{\a\b} \Big)^{1/2}
=\ft12 a Q \sqrt{\a(N+1-\a)} \ ,\label{suncharge}
\ee
where $a$ is given $a^2=\Delta=24/(N(N+1)(N+2))$.  Under these circumstances 
the solutions reduce to single-scalar solutions, given by
\bea ds^2 &=& \eta_{\mu\nu}\, dx^\mu\, dx^\nu + H^{1/\Delta}\, (dr^2 +
r^2\, d\theta^2) \ ,\nn\\
e^{a\phi/2} &=& H = 1 + k\, \log r\ ,\qquad \chi=4\, Q\, \theta\ ,
\eea
and have mass
\be
m=\fft{2Q}{a}\ .
\ee
It is easy to verify that this is always larger than the total mass of
the widely-separated constituents, $m_\infty=\sum_\a Q_\a$.  The calculation
of the eigenvalues of the \bog matrix becomes increasingly complicated with
increasing $N$.  For example, for the $SL(5,R)$ case we find
\be
\mu=m \pm \sqrt{Q_1^2 + Q_2^2 + Q_3^2 + Q_4^2 \pm 2
\sqrt{(Q_1 Q_3)^2 + (Q_1 Q_4)^2 + (Q_2 Q_4)^2}}\ ,
\ee
whilst for $SL(6,R)$ we find that $\mu=m\pm \kappa$, where $\kappa$ denotes
the roots of the quartic equation $\kappa^4 -2 \kappa^2\, \a -8 \kappa \, Q_1
\, Q_3 \, Q_5 + \b=0$, and 
\bea
\a&=&Q_1^2 + Q_2^2 + Q_3^2 + Q_4^2 + Q_5^2\ ,\\
\b&=& \a^2 -4\Big( (Q_1 Q_3)^2 + (Q_1 Q_4)^2 +  (Q_1 Q_5)^2 +
          (Q_2 Q_4)^2 +  (Q_2 Q_5)^2 +  (Q_3 Q_5)^2 \Big)\ .\nonumber
\eea
For all $N$, the solutions are non-supersymmetric for generic values of the
charges.  However, they can be reduced to the previously-known supersymmetric
solutions if appropriate charges are set to zero, such that the remaining
charges $Q_\a$ have non-adjacent indices.  In these cases, the solutions
preserve $2^{-n}$ of the supersymmetry, where $n$ is the number of charges
remaining.

\subsection{Fission and fusion bound states of $p$-brane solitons}

       So far, we have obtained a large class of $p$-branes solutions
in the toroidally-compactified M-theory.  It is useful to classify
and organise these solutions.  One approach is to observe that supersymmetric
$p$-branes that carry a single electric or magnetic charge, and hence
preserve half the supersymmetry, can be interpreted as the constituents
from which all the multiply-charged
$p$-branes can be constructed as bound states.  The binding energy can
be zero, positive and negative, depending on the specific choice of
constituents.

      The binding energy of a $p$-brane can easily be calculated by
comparing its mass with the sum of the masses of its individual
constituents when their locations are widely separated.  Of course if
the binding energy is non-zero, this configuration will not be an
exact solution. However, it can be made arbitrarily good by taking the
separations to be sufficiently large.  Since each individual
constituent satisfies the Bogomol'nyi bound, {\it i.e.}\ its mass is
equal to the charge, it follows that the total mass when the
constituents are widely separated is given by
\be
m_\infty = \sum_\a^N Q_\a\ ,\label{minfinity}
\ee
where $Q_\a$ are the charges of the individual constituents.

      There are various ways to obtain multiply-charged $p$-branes.
The simplest way is to act with a
U-duality rotation on a singly-charged $p$-brane.  The new solution preserves
the same fraction of supersymmetry as the original one.  Since these solutions
contain multiple charges, they can be viewed as bound states. Such bound
states usually have positive binding energy.  For example, the 8-dimensional
dyonic membrane can be obtained by an $SL(2,Z)$  T-duality transformation
from a purely electric or purely magnetic membrane, as we discussed in section
4.4.   The mass of the dyonic membrane is given by (\ref{d8dyonmass}), which
is always smaller than the combined mass of the two widely-separated
electrically-charged and magnetically-charged membranes.  Thus the dyonic
membrane is a bound state of these two basic constituents with positive
binding energy.   Another simple example is provided by the two string
solutions of type IIB supergravity in $D=10$.  One of these uses the NS-NS
3-form field strength, whilst the other uses the R-R 3-form.  There is a
non-perturbative $SL(2,Z)$ symmetry of the type IIB theory, which
rotates between the NS-NS string and the R-R string solutions. Thus
one obtains a bound state of the NS-NS string and the R-R string by
acting with an $SL(2,Z)$ transformation on either the NS-NS or the R-R
string solution \cite{sch}.  The mass of the bound state is given by
$m=\sqrt{Q_{\rm NS-NS}^2 + Q_{\rm R-R}^2}$, and hence it has positive
binding energy \cite{w2}.  In general, all the bound states that are
obtained by acting with U-duality rotations on singly-charged
$p$-branes have positive binding energy.

        The multiply-charged solutions we obtained in section 4.1 are
of a different type.  They are not related to the single-charge solutions of
any of their constituents by U-duality rotations.  One way to see this is that
they preserve a different fraction of the supersymmetry.  The masses of these
solutions are given by (\ref{massform}) and hence they can be
viewed as bound states with zero binding energy \cite{r1,kklp,drbound}.  
For example, the
$a=1$, $1/\sqrt3$ and 0 black holes in $D=4$ can be viewed as bound
states of two, three or four $a=\sqrt3$ black holes \cite{r1}.  Another
example is provided by the dyonic string in $D=6$, discussed in
section 4.4, which can be viewed as a bound state of an electric and
a magnetic string \cite{kklp}.  These bound states are in general
supersymmetric, although non-supersymmetric solutions can also arise.  For
example, the $a=0$ Reissner-Nordstr{\o}m back hole in $D=4$ can be
non-supersymmetric for certain choices of sign of the four
constituent charges, which appear quadratically in the equations of motion,
but linearly in the supersymmetry transformation rules
\cite{lpmulti,ko}, as discussed in section 4.2.   For all these solutions, 
strictly speaking, the term bound state is a misnomer since the
binding energy is actually zero.   This zero binding energy is
consistent with the fact that the charges in the above multi-charge
solutions can be located independently; the bound states can be
``pulled apart'' into constituents that can sit in static equilibrium
at any separation.  This can be seen from the fact that
generalisations of the solution (\ref{gensol2}) exist where the functions
$H_\a$ can be any harmonic functions
\cite{t1} on the transverse space $y^m$, implying in particular that each
individual charge can be located at any point in the transverse space. 
We shall discuss such multi-centre solutions further in section 5.1.  This
type of multi-charge, multi-centre solution was first discussed in
\cite{klopp}, where a four-charge supersymmetric black hole in $D=4$ was
``split'' into two $a=1$ two-charge black holes (which themselves can be
further split into $a=\sqrt3$ black holes).

        The third type of bound states are those with negative binding
energy.  These solutions are non-supersymmetric.  Examples are
provided by the 4-dimensional dyonic black hole, and by the $SL(N+1, R)$
solitons in various dimensions, which were discussed in section 4.4 and 4.5
respectively.   It is easy to see from the mass formulae for the
solutions that these bound states have negative binding energy.  Note
that the dyonic black hole in section 4.4 reduces to the 4-dimensional
Reissner-Nordstr{\o}m black hole when the electric and magnetic charges
are equal.  Thus a Reissner-Nordstr{\o}m black hole in $D=$ can be viewed
either as an inert bound state of four constituents with zero binding
energy, or as a  ``dyonic fission bomb'' of negative binding energy,
comprised of an electric black hole and a magnetic black hole of equal
charges.  The two cases are distinguished by the choice of field strengths
that carry the charges.

          Finally, let us note that in $D=8$ the only dyonic membrane is
the one that can be obtained by acting on the purely electric or
purely magnetic membrane with a T-duality
$SL(2,R)$ transformation \cite{ilpt}.  The resulting dyonic membrane preserves
half of the supersymmetry, and is a bound state with positive binding
energy.  In
$D=6$ and $D=4$, rotations of this kind cannot be used to convert a solution
with purely electric or purely magnetic charges into a dyonic solution; other
field strengths will also acquire charges at the same time. In $D=6$ and
$D=4$, dyonic solutions can be constructed directly, as we saw in
section 4.4.   In $D=6$, the dyonic string preserves $\ft14$ of the
supersymmetry, and is a bound state with zero binding energy.  In
$D=4$, the dyonic black hole solution is non-supersymmetric, and is a bound
state with negative binding energy.
\vfill\eject

\section{Dimensional reduction and oxidation}

     In the previous sections, we have extensively discussed classes of
$p$-brane solitons that arise as solutions in the toroidal
compactifications of M-theory.  Of course since the toroidally-compactified
supergravities are themselves consistent truncations of $D=11$
supergravity, it follows that if  higher-dimensional $p$-brane solutions
are themselves dimensionally reduced, they will give rise to solutions of
the lower-dimensional theories.  In fact many, but not all, of the
lower-dimensional $p$-brane solitons can be obtained simply as the dimensional
reductions of $p$-branes in higher dimensions.  

     As we shall
discuss below, there are two ways of dimensionally reducing a $p$-brane
solution to give another such solution in one lower dimension.  The simpler
of the two is ``diagonal'' reduction, in which one of the spatial
coordinates on the $p$-brane world-volume is used for the Kaluza-Klein
reduction.  This removes one dimension from the spacetime and from the
world-volume simultaneously, and thus we go from a $p$-brane in $D$
dimensions to a $(p-1)$-brane in $D-1$ dimensions, \ie $(D,p)\rightarrow
(D-1,p-1)$.  The other procedure is known as ``vertical'' dimensional
reduction, and in this case one of the transverse-space coordinates is used
for the Kaluza-Klein reduction, taking us from $(D,p)$ to $(D-1,p)$.  This
is more complicated to implement, because to perform a Kaluza-Klein
reduction on a coordinate it is necessary that translations along that
direction should be a Killing symmetry, whereas a standard single $p$-brane
soliton depends isotropically on all the coordinates $y^m$ of the Cartesian
transverse space. It is first necessary to construct a multi-centre
$p$-brane solution in the higher dimension, with the centres periodically
aligned along the chosen coordinate axis, such that as the continuum limit
is taken, the solution becomes independent of this coordinate.  By means of
these two reduction procedures, a given
$p$-brane in $D$ dimensions gives rise to a $(p-1)$-brane and a $p$-brane
in $D-1$ dimensions.

   However, whilst it is certainly true,
owing to the consistency of the truncations to the lower-dimensional
supergravities, that all the lower-dimensional $p$-branes will also be
solutions in the higher-dimensional theories, not all of them ``oxidise''
back to simple $p$-brane solutions in the higher dimensions.  Consequently,
there will be other types of solutions of the higher-dimensional
supergravities that are not immediately recognisable as $p$-branes, which
nevertheless, upon dimensional reduction, give rise to lower-dimensional
$p$-brane solitons of the usual kind.  These more complicated
higher-dimensional configurations can be interpreted as intersections of
various $p$-branes, or as ``twisted'' $p$-branes.  We shall discuss them
further in section 5.2 below, having first described the simpler situation
of the diagonal and vertical reduction of $p$-branes.

\subsection{Diagonal and vertical dimensional reduction}

     First, let us consider the diagonal dimensional reduction of an
$N$-charge $p$-brane solution in $D$ dimensions to a $(p-1)$-brane in $D-1$
dimensions.  Thus we begin with the metric (\ref{gensol2}),
\be
ds_{\sst D}^2=\prod_{\a=1}^{N} H_\a^{-\ft{\tilde d}{(D-2)}}\, dx^\mu dx^\nu
\eta_{\mu\nu} +
\prod_{\a=1}^N H_\a^{\ft{d}{(D-2)}}\, (dr^2 + r^2 d\Omega^2) \ ,\label{dsol}
\ee
where the $H_\a$ are harmonic functions on the transverse space, of the
form $H_\a=1+ \ft{\lambda_\a}{\td d} r^{-\td d}$.  The Kaluza-Klein reduction
of this solution to $D-1$ dimensions is therefore described by the metric
$ds_{\sst D-1}^2$, related to $ds_{\sst D}^2$ by 
\be
ds_{\sst D}^2 = e^{2\a\varphi}\, ds_{\sst D-1}^2 + e^{-2(D-3)\a\varphi} \,
dz^2\ ,\label{kkred}
\ee
where $\a=(2(D-2)(D-3))^{-1/2}$ and the coordinate $z$ is one of the spatial
coordinates $x^i$ in the $D$-dimensional $p$-brane world-volume.  Thus we
see that $e^{2(D-3)\a\varphi}=\prod_\a H_\a^{\td d/(D-2)}$, and hence the
$(D-1)$-dimensional metric is given by
\be
ds_{\sst D-1}^2 = \prod_{\a=1}^{N} H_\a^{-\ft{\tilde d}{(D-3)}}\, dx^\mu
dx^\nu\eta_{\mu\nu} +
\prod_{\a=1}^N H_\a^{\ft{d-1}{(D-3)}}\, (dr^2 + r^2 d\Omega^2) \ ,
\label{dmsol}
\ee
where now the $\mu$ index ranges over one fewer spatial indices than in
(\ref{dsol}).  We see that (\ref{dmsol}) is describing an $N$-charge
$(p-1)$-brane solution in $D-1$ dimensions, with otherwise precisely the
same structural form as (\ref{dsol}).  Although we have now acquired one
more dilatonic scalar, namely $\varphi$, it is evident from the solutions
for the dilatons $\varphi_\a$ in (\ref{gensol2}) that a certain linear
combination of $\varphi_\a$ and $\varphi$ vanishes, and hence there is no
net increase in the number of excited scalars.  In fact a careful
calculation shows that the excited dilatonic degrees of freedom in the
dimensionally-reduced solution are precisely of the same form as in the
$D$-dimensional solution, where the appropriate changes are made to account
for the additional component to the dilaton vectors that is acquired by
virtue of the reduction procedure.  Note, incidentally, that this
description of the diagonal dimensional reduction can easily be extended
also to the case of non-extremal $p$-brane solutions.

    Now let us consider the vertical dimensional reduction process.  To do
this, we need first to construct more general multi-centre $p$-brane
solutions in $D$ dimensions.  In fact the way in which the single-centre
solutions (\ref{gensol2}) are written already suggests the form of the more
general solutions. It is quite straightforward to show that the
harmonic functions $H_\a$ need not be restricted to be single-centre
isotropic functions, and that we still have solutions of the
equations if they are taken to have the quite general multi-centre form
\be
H_\a = 1 + \sum_i \fft{k_i^\a}{|\vec y -\vec y_i^\a|^{\td d}}\ ,
\ee
where we now use the Cartesian coordinates $y^m$ on the $(D-d)$-dimensional 
transverse space,
and $k_i^\a$ are arbitrary constants.  As a special case we may choose the
centres to be distributed periodically along a particular axis, which for 
simplicity could be the last of the $y^m$ coordinates, such that in the 
continuum limit the harmonic functions become independent of this coordinate.
The integration over the continuous line of centres reduces the powers in the
denominators in the harmonic functions, $\int_{-\infty}^\infty dz (r^2+z^2)^{
-\td d/2} \sim r^{-\td d+1}$, with the net result that what remains is 
harmonic with respect to a transverse space of one lower dimension. 
(Multi-centre extremal static solutions have been constructed in
\cite{ma,pa,gh}, and their application for dimensional reduction was
considered in \cite{k2,ghl}.  A 
detailed discussion of this, including the cases where the resulting
harmonic functions have logarithmic, or even linear, coordinate dependence,
can be found in \cite{lpsvert}.)  Reading off from (\ref{kkred}), 
where now the reduction coordinate is the last of the $y^m$ variables in $D$
dimensions, we see that this time $e^{-2(D-3)\a\varphi} = \prod_\a 
H_\a^{d/(D-2)}$, where the $H_\a$ are harmonic in the remaining 
$(D-d-1)$-dimensional transverse space.  Substituting back into (\ref{kkred}),
we find that the resulting metric in $D-1$ dimensions has the form
\be
ds_{\sst D-1}^2 = \prod_{\a=1}^{N} H_\a^{-\ft{(\tilde d-1)}{(D-3)}}\, dx^\mu
dx^\nu\eta_{\mu\nu} +
\prod_{\a=1}^N H_\a^{\ft{d}{(D-3)}}\, dy^m\, dy^m \ .
\label{dmmsol}
\ee
This is precisely of the same structural form as the original $p$-brane
metric in the higher dimension, except that here the dimension of spacetime
has been lowered by removing one of the transverse dimensions, while
preserving the dimension of the worldvolume.

\subsection{Bound states as intersecting $p$-branes}

      In section 4, we obtained large classes of $p$-brane solutions.  In
section 4.6, we discussed a way to classify some of them, by viewing 
multiply-charged $p$-branes as bound states of singly-charged
$p$-branes, which preserve half the supersymmetry. Another approach is to
organise the various solutions using the U-duality group of the
theory. In particular, it has been shown that $p$-brane solutions form
representations of the Weyl group of the U-duality group \cite{lpsweyl}. 
Since duality symmetry commutes with supersymmetry, each member of
such a U-duality multiplet preserves the same fraction of the supersymmetry.

      Both of these organisational schemes are applicable in a given, 
fixed, dimension.  A different approach is to interpret lower-dimensional
solutions from the viewpoint of the fundamental dimension of the
theory, namely $D=11$ in the case of M-theory.  In other words, the
lower-dimensional solutions can be oxidised, by the inverse of the
Kaluza-Klein reduction procedure, to solutions in $D=11$.  All maximal
supergravities in the lower dimensions can be obtained by Kaluza-Klein
reduction of $D=11$ supergravity by truncating out the massive modes.
Since such a truncation is consistent, it implies that any lower
dimensional solution can be oxidised back to a solution of $D=11$
supergravity. It has been shown that lower-dimensional supersymmetric
$p$-branes can be viewed as intersecting M-branes, or boosted or
twisted intersecting M-branes in $D=11$ [51,54-58].  We shall present
a few examples to illustrate this.

         In the previous subsection, we saw that a $p$-brane can
undergo both vertical and diagonal dimensional reduction.  Thus, for 
example, a membrane in $D=11$ can become a string in $D=10$ by diagonal
dimensional reduction, and then by four steps of vertical reduction it
becomes an electrically-charged string in $D=6$.  On the other hand,
the 5-brane in $D=11$ can be vertically reduced to a 5-brane in
$D=10$, and then diagonally reduced to give a magnetically-charged
string in $D=6$.  Thus electric strings and magnetic strings in $D=6$
can be oxidised back to give membranes and 5-branes respectively in
$D=11$.  Naturally a dyonic string $D=6$, which carries both electric
and magnetic charges, can also be oxidised to $D=11$, where it becomes
an intersection a membrane and a 5-brane in $D=11$.  The metric of the
$D=11$ solution, obtained by simply reversing the Kaluza-Klein
reduction process described by (\ref{metred}), is given by
\bea
ds_{11}^2 &=& H_e^{-\ft23} H_m^{-\ft13} (-dt^2 + dx_1^2) + H_e^{-\ft23} 
          H_m^{\ft23} dz_1^2 + H_e^{\ft13} H_m^{\ft23} dy^m dy^m
\nonumber\\
&& +  H_e^{\ft13} H_m^{-\ft13}(dz_2^2 + dz_3^2 + dz_4^2 + dz_5^2)
\ ,\label{intersecting25} 
\eea
where $H_e$ and $H_m$ are harmonic functions in the 4-dimensional
space $y^m$, associated with the electric and magnetic charges respectively,
as in (\ref{multfsol}). 
This solution gives rise to a dyonic string in $D=6$ when compactified on
a 5-torus with coordinates $z_i$, $i-1, \ldots, 5$.  If the magnetic
charge of the dyonic string is set to zero, in which case $H_m=1$, the
solution becomes a configuration describing membranes with world
volume coordinates $(t, x_1, z_1)$, whose charges are uniformly
distributed over the hyperplane $(z_2, z_3, z_4, z_5)$.  On the other
hand, if instead the electric charge is set to zero, in which case $H_e=1$, it
describes a line (along the $z_1$ axis) of uniformly distributed 5-branes,
with world volume coordinates $(t, x_1, z_2, z_3, z_4, z_5)$.   Thus the
solution (\ref{intersecting25}) describes an interpolation between membranes
and 5-branes, and hence is called an intersection of membranes and 5-branes.

       Another example is provided by the supersymmetric
Reissner-Nordstr{\o}m black hole in $D=4$.  It is a bound state of
four basic $a=\sqrt3$ constituent black holes.  There are various ways
of constructing such 4-charge black holes, the set of which form a
representation under the Weyl group of $E_7$ \cite{lpsweyl}.  One member of
such a multiplet is the solution involving the field strengths $F_2^{(12)}$,
$F_2^{(34)}$  carrying electric charges, and $F_2^{(13)}$ and $F_2^{(24)}$
carrying magnetic charges.  Oxidising to $D=11$, the metric of this
particular 4-charge black hole solution becomes
\bea
ds_{11}^2 &=& \Big(\fft{H_3H_4}{H_1H_2}\Big)^{\ft16}\, ds_4^2 +
     \Big(\fft{H_2 H_3^2}{H_1^2 H_4} \Big)^{\ft13} dz_1^2 +
     \Big(\fft{H_2 H_4^2}{H_1^2 H_3} \Big)^{\ft13} dz_2^2 +
     \Big(\fft{H_1 H_3^2}{H_2^2 H_4} \Big)^{\ft13} dz_3^2\nonumber\\
&&  + \Big(\fft{H_1 H_4^2}{H_2^2 H_3} \Big)^{\ft13} dz_4^2 +
     \Big(\fft{H_1 H_2^2}{H_3^2 H_4} \Big)^{\ft13} (dz_5^2
                    + dz_6^2 + dz_7^2)\ ,\label{d11rn}
\eea
where $ds_4^2$ is the metric for the 4-charge black hole, given by
\be
ds_4^2 =-( H_1H_2H_3H_4)^{-\ft12}dt^2 + (H_1H_2H_3H_4)^{\ft12} dy^m
dy^m\ .\label{rn}
\ee
The functions $H_\a$ are harmonic in the 3-dimensional 
transverse space described by the coordinates $y^m$, and the metric (\ref{rn})
becomes that of the usual extremal Reissner-Nordstr{\o}m black hole if all
the charges, and hence all the $H_\a$, are equal.   Thus it is easy to see
now with the explicit metric (\ref{d11rn}) that this Reissner-Nordstr{\o}m
black hole becomes an intersection of two membranes and two 5-branes in
$D=11$.

     The above two examples have the feature that the field
strengths that are involved in the solutions all come from the
dimensional reduction of the 4-form field strength in $D=11$.  The
oxidation of these bound-state $p$-branes in lower dimensions 
give intersections of M-branes in $D=11$, with the electrically-charged
constituents becoming membranes, and the magnetically-charged constituents
becoming 5-branes.  The complete classification of such intersecting M-branes
can be found in \cite{bdejs}. Of course in lower dimensions 
field strengths can also come from the reduction of the 11-dimensional metric
tensor.  In these cases, the oxidised metric in $D=11$ will acquire
off-diagonal components.  An electrically-charged constituent in the
lower dimension will describe a ``boost'' in $D=11$,
involving an off-diagonal component that mixes time and spatial directions 
in the world-volume of the M-brane.  On the other hand, a
magnetically-charged constituent will oxidise to a ``twisted'' metric in
$D=11$, involving a monopole-like configuration in the transverse space of
the M-brane.

         So far we described how bound states with zero binding
energy can be viewed as intersections of M-branes (with possible
boosts or twists) in $D=11$.  As we saw in section 4.6, bound states can also
exist that have either positive or negative binding energy. 
Bound-state $p$-branes with positive binding energy are usually obtained
from U-duality rotations of singly-charged constituent $p$-branes.  The
rotation involves non-linear transformations of axions,
which are the dimensional reduction from $D=11$ of the 3-form potential
or the metric.  Such a solution has a complicated metric
structure in $D=11$, and cannot be simply interpreted as an intersection of
M-branes.  However, these solutions are in the same U-duality multiplet as
simple solutions that can be interpreted as intersecting M-branes. 

       Bound states with negative binding energy can be oxidised
straightforwardly, as in the case of the supersymmetric bound states
with zero binding energy.  We may take the dyonic  black hole
solution in $D=4$ as an illustration.  Again, we shall only consider
the case where the 2-form field strength involved in the solution
comes from the 4-form in $D=11$.  For example, let us consider the
dyonic black hole where the field strength $F_2^{(12)}$carries the electric
and magnetic charges.  The solution is given in section 4.4, and upon
oxidation to $D=11$ the metric becomes
\bea
ds_{11}^2 &=& (\fft{T_e}{T_m})^{-\ft16} ds_4^2 +
              (\fft{T_e}{T_m})^{-\ft23} (dz_1^2 + dz_2^2) +
              (\fft{T_e}{T_m})^{\ft13} (dz_3^2 + \cdots + dz_7^2)\ ,
\label{neginter}
\eea
where $ds_4^2$, $T_e$ and $T_m$ are given by (\ref{d4dyonmetric}) and
(\ref{dyonsol}).  Thus when the electric charge $Q_e$ or the magnetic
charge $Q_m$ vanishes, the metric reduces to a 5-brane or a membrane
respectively.  Note that both (\ref{d11rn}) and (\ref{neginter}) give
rise to Reissner-Nordstr{\o}m black holes in $D=4$; however the
former has zero binding energy whilst the latter has negative binding
energy.  The main difference from the 11-dimensional point of view is that
the latter gives rise to a metric with a larger group of symmetries. 
We see from this example that a membrane can intersect with a 5-brane in
different ways in $D=11$.  In one form of intersection there is zero 
binding energy and the solution reduces to the dyonic string in
$D=6$, while in another there is negative binding
energy, and the solution reduces to the dyonic black hole in
$D=4$.  A third possibility is for the intersection of the membrane and
5-brane to have positive binding energy, reducing to the dyonic membrane
in $D=8$.

\pagebreak


\begin{thebibliography}{99}

\bm{nam} Y. Nambu, {\sl Quark model and the factorisation of the Veneziano
amplitude}, in Symmetries and Quark Models (Ed. R. Chand, Gordon and Breach).

\bm{yon} T. Yoneya, {\sl Quantum gravity and the zero-slope limit of
the generalised Virasoro model}, Nuovo Cim. Lett. {\bf 8} (1973) 951.

\bm{scsc} J. Scherk and J.H. Schwarz, {\sl Dual models and the 
geometry of spacetime}, Phys. Lett. {\bf B52} (1974) 347.

\bm{grsc} M.B. Green and J.H. Schwarz, {\sl Anomaly cancellation in 
supersymmetric $D=10$ gauge theory and superstring theory}, Phys. Lett.
{\bf B149} (1984) 117.

\bm{ht} C.M. Hull and P.K. Townsend, {\sl Unity of superstring dualities}, 
Nucl. Phys. {\bf B438} (1995) 109.

\bm{w1} E. Witten, {\sl String theory dynamics in various dimensions},
Nucl. Phys. {\bf B443} (1995) 184.

\bm{cjs} E. Cremmer, B, Julia and J. Scherk, {\sl Supergravity theory in 
eleven dimensions}, Phys. Lett. {\bf B76} (1978) 409.

\bm{dnp} M.J. Duff, B.E.W. Nilsson and C.N. Pope, {\sl Kaluza-Klein
supergravity}, Phys. Reps. {\bf 130} (1986) 1.

\bm{hw} P. Horava and E. Witten, {\sl Heterotic and type I string 
dynamics from eleven-dimensions}, Nucl. Phys. {\bf B460} (1996) 506.

\bm{lpsol} H. L\"u and C.N. Pope, {\sl $p$-brane solitons in maximal 
supergravities}, Nucl. Phys. {\bf B465} (1996) 127.

\bm{ss1} J. Scherk and J.H. Schwarz, {\sl How to get masses from extra 
dimensions}, Nucl. Phys. {\bf B153} (1979) 61.

\bm{bdgpt} E. Bergshoeff, M. de Roo, M.B. Green, G. Papadopoulos and
P.K. Townsend, {\sl Duality of type II 7-branes and 8-branes}, Nucl. Phys. 
{\bf B470} (1996) 113.

\bm{clpst} P.M. Cowdall, H. L\"u, C.N. Pope, K.S. Stelle and P.K. Townsend,
{\sl Domain walls in massive supergravities}, hep-th/9608173, to appear
in Nucl. Phys. {\bf B}.

\bm{lpdomain} H. L\"u and C.N. Pope, {\sl Domain walls from M-branes},
hep-th/9611079.

\bm{llp} I.V. Lavrinenko, H. L\"u and C.N. Pope, {\sl From topology to
generalised dimensional reduction}, hep-th/9611134, to appear in Nucl. Phys.
{\bf B}.

\bibitem{dghr}A. Dabholkar, G.W. Gibbons, J.A. Harvey and
F. Ruiz Ruiz, {\sl Superstrings and solitons}, 
Nucl. Phys. {\bf B340} (1990) 33.

\bm{str} A. Strominger, {\sl Heterotic solitons}, Nucl. Phys.
{\bf B343} (1990) 167.

\bm{ds5brane} M.J. Duff and K.S. Stelle, {\sl Multi-membrane solutions
of $D=11$ supergravity}, Phys. Lett. {\bf B253} (1991) 113.

\bm{dl1} M.J. Duff and J.X. Lu, {\sl Elementary fivebrane solutions of
$D=10$ supergravity}, Nucl. Phys. {\bf B354} (1991) 141;
{\sl Strings from fivebranes}, Phys. Rev. Lett. {\bf 66} (1991) 1402;
{\sl The selfdual type IIB superthreebrane}, Phys. Lett. {\bf B273}
(1991) 409.

\bm{chs} C.G. Callan, J.A. Harvey and A. Strominger, {\sl World sheet
approach to heterotic instantons and solitons}, Nucl. Phys. 
{\bf B359} (1991) 611; {\sl Worldbrane actions for string solitons},
Nucl. Phys. {\bf B367} (1991) 60.

\bm{guv} R. G\"uven, {\sl Black $p$-brane solutions of $D=11$
supergravity theory}, Phys. Lett. {\bf B276} (1992) 49.

\bm{k1} R.R. Khuri, {\sl A comment on the stability of string
monopoles},  Phys. Lett. {\bf B307} (1993) 302;
{\sl Classical dynamics of macroscopic strings}, Nucl. Phys. 
{\bf B403} (1993) 335.

\bm{cy} M. Cvetic and D. Youm, {\sl Static four-dimensional abelian
black holes in Kaluza-Klein theory}, Phys. Rev. {\bf D52}, (1995)
2144; {\sl Kaluza-Klein black holes within heterotic string theory on
a Torus}, Phys. Rev. {\bf D52} (1995) 2574; {\sl All the
four-dimensional static, spherically symmetric solutions of abelian
Kaluza-Klein theory}, Phys. Rev. Lett. {\bf B75} (1995) 4165.

\bm{dklreview} M.J. Duff, R.R. Khuri and J.X. Lu, {\sl String
solitons}, Phys. Rep. {\bf 259} (1995) 213.

\bm{lpss1} H. L\"u, C.N. Pope, E. Sezgin and K.S. Stelle, {\sl Stainless
super $p$-branes}, Nucl. Phys. {\bf B456} (1995) 669.

\bm{page} D.N. Page, {\sl Classical stability of round and squashed 
seven spheres in eleven-dimensional supergravity}, Phys. Rev.
{\bf D28} (1983) 2976.

\bm{dlps} M.J. Duff, H. L\"u, C.N. Pope and E. Sezgin, {\sl
Supermembranes with fewer supersymmetries}, Phys. Lett. {\bf B371}
(1996) 206. 

\bm{lpmulti} H. L\"u and C.N. Pope, {\sl Multi-scalar $p$-brane solitons},
Int. J. Mod. Phys. {\bf A12} (1997) 437.

\bm{kklp} N. Khviengia, Z. Khviengia, H. L\"u and C.N. Pope, {\sl
Intersecting M-branes and bound states}, Phys. Lett. {\bf B388}
(1996) 21.

\bm{drbound} M.J. Duff and J. Rahmfeld, {\sl Bound states of black holes 
and other p-branes}, Nucl. Phys. {\bf B481} (1996) 332.

\bm{lmp2} H. L\"u, S. Mukherji and C.N. Pope, {\sl From $p$-branes
to cosmology}, hep-th/9612224.

\bm{dlpblack} M.J. Duff, H. L\"u and C.N. Pope, {\sl The black branes
of M-theory}, Phys. Lett. {\bf B382} (1996) 73.

\bm{jxlu} J.X. Lu, {\sl ADM mass for black strings and $p$-branes},
Phys. Lett. {\bf B313} (1993) 29.

\bm{lpxtoda} H. L\"u, C.N. Pope and K.W. Xu, {\sl Liouville and Toda solitons
in M-theory}, Mod. Phys. Lett. {\bf A11} (1996) 1785.

\bm{dfkr} M.J. Duff, S. Ferrara, R.R. Khuri and J. Rahmfeld, {\sl
Supersymmetry and dual string solitons}, Phys. Lett. {\bf B356}
(1995) 479.

\bm{gk} G.W. Gibbons and R.E. Kallosh, {\sl Topology, entropy and Witten
index of dilaton black holes}, Phys. Rev. {\bf D51} (1995) 2839.

\bm{ilpt} J.M. Izquierdo, N.D. Lambert, G. Papadopoulos and P.K.
Townsend, {\sl Dyonic membrane}, Nucl. Phys. {\bf B460} (1996) 560.

\bm{lpsln} H. L\"u and C.N. Pope, {\sl $SL(N+1, R)$ Toda solitons in 
supergravities}, hep-th/9607027.

\bm{ls} A.N. Leznov and M.V. Saveliev, {\sl Group-theoretical methods
for integration of nonlinear dynamical systems}, Basel: Birkhauser 1992.

\bm{a} A. Anderson, {\sl An elegant solution of the $n$-body Toda problem},
J. Math. Phys. {\bf 37} (1996) 1349.

\bm{sch} J.M. Schwarz, {\sl An $SL(2,Z)$ multiplet of type IIB
superstrings} Phys. Lett. {\bf B360} (1995) 13.

\bm{w2} E. Witten, {\sl Bound states of strings and $p$-branes},
Nucl. Phys. {\bf B460} (1996) 335.

\bm{r1} J. Rahmfeld, {\sl Extremal black holes as bound states}, 
Phys. Lett. {\bf B372} (1996) 198.

\bm{ko} R.R. Khuri and T. Ortin, {\sl A non-supersymmetric dyonic extreme
Reissner-Nordstr{\o}m black hole}, Phys. Lett. {\bf B373} (1996) 56.

\bm{ma} D. Majumdar, Phys. Rev. {\bf 72} (1947) 390.

\bm{pa} A. Papapetrou, Proc. R. Irish Acad. {\bf A51} (1947) 191.

\bm{gh} G.W. Gibbons and S.W. Hawking, {\sl Gravitational
multi-instantons}, Phys. Lett. {\bf B78} 430.

\bm{k2} R. Khuri, {\sl A heterotic multi-monopole solution }, 
Nucl. Phys. {\bf B387} (1992) 315.

\bm{ghl} J.P. Gauntlett, J.A. Harvey and J.L. Liu, {\sl Magnetic
monopoles in string theory}, Nucl. Phys. {\bf B409} (1993) 363.

\bm{lpsvert} H. L\"u, C.N. Pope and K.S. Stelle, {\sl Vertical versus
diagonal dimensional reduction for $p$-branes}, Nucl. Phys. {\bf B481}
(1996) 313.

\bm{t1} A.A. Tseytlin, {\sl Harmonic superpositions of M-theory},
Nucl. Phys. {\bf B475} (1996) 149.

\bm{klopp} R.R. Kallosh, A. Linde, T. Ortin, A. Peet and A. van Proeyen,
{\sl Supersymmetry as a cosmic censor}, Phys. Rev. {\bf D46} (1992) 5278.

\bm{lpsweyl} H. L\"u, C.N. Pope and K.S. Stelle, {\sl Weyl group invariance
and $p$-brane multiplets}, Nucl. Phys. {\bf B476} (1996) 89.

\bm{pt} G. Papadopoulos and P.K. Townsend, {\sl Intersecting
M-branes}, Phys. Lett. {\bf B380} (1996) 273.

\bm{kt} I.R. Klebanov and A.A. Tseytlin, {\sl Intersecting M-branes as four
dimensional black holes}, Nucl. Phys. {\bf B475} (1996) 179.

\bm{gkt} J.P. Gauntlett, D.A. Kastor and J. Traschen, {\sl Overlapping
branes in M-theory}, Nucl. Phys. {\bf B478} (1996) 544.

\bm{bl} V. Balasubramanian and F. Larsen, {\sl On D-branes and black holes
in four dimensions}, Nucl. Phys. {\bf B478} (1996) 199.

\bm{bdejs} E. Bergshoeff, M. de Roo, E. Eyras, B. Janssen, J.P. van
der Schaar, {\sl Multiple intersections of D-branes and M-branes},
hep-th/912095.







\end{thebibliography}
\end{document}